\documentclass[traditabstract]{aa} % for the abstract without structuration
\usepackage{floatrow}
\usepackage{natbib}
\usepackage{amsmath}
\usepackage{array}
\usepackage{booktabs}
\usepackage{cellspace}

\usepackage{amssymb}
\restylefloat{table}
\usepackage{hyperref}

\usepackage{xcolor}
\definecolor{dark-red}{rgb}{0.9,0.0,0.0}
\definecolor{dark-blue}{rgb}{0.15,0.15,0.9}
\definecolor{dark-green}{rgb}{0.15,0.8,0.15}
\definecolor{medium-blue}{rgb}{0,0,0.9}
\hypersetup{
    colorlinks, linkcolor=red,
    citecolor={dark-blue} , urlcolor={medium-blue}
}

\usepackage{lscape}
\usepackage{graphicx}
\usepackage[varg]{txfonts}
\usepackage{blindtext}
\usepackage{lipsum}

\usepackage{changepage} % for \begin{adjustwidth}{-Xcm}{} \end{adjustwidth}

\begin{document}
   \title{Precise radial velocities of giant stars \\
   }
\subtitle{XI. Two brown dwarfs in 6:1 mean motion resonance \\ around the K giant star
$\nu$\,Ophiuchi.\thanks{Based on observations collected at Lick Observatory, University of
California and on observations collected at the European Southern Observatory, Chile, under program IDs 088.D-0132, 089.D-0186, 090.D-0155 and 091.D-0365.}}

    \author{Andreas Quirrenbach\inst{1}, Trifon Trifonov\inst{1}$^{,}$\inst{2}$^{,}$\inst{3}, Man Hoi Lee\inst{2}$^{,}$\inst{4},
    \and Sabine Reffert\inst{1}
    }

   \institute{ Landessternwarte, Zentrum f\"ur Astronomie der Universit\"at Heidelberg, K\"{o}nigstuhl 12, 69117 Heidelberg, Germany\\ \vspace{-3mm}
               \and
   Department of Earth Sciences, The University of Hong Kong, Pokfulam Road, Hong Kong\\ \vspace{-3mm}
              \and
   Max-Planck-Institut f\"ur Astronomie, K\"onigstuhl 17, 69117 Heidelberg, Germany\\ \vspace{-3mm}
              \and
   Department of Physics, The University of Hong Kong, Pokfulam Road, Hong Kong\\
}

   \date{Received 13 October 2018 / Accepted 9 January 2019}

% \abstract{}{}{}{}{}
% 5 {} token are mandatory

  \abstract
   {
We present radial-velocity (RV) measurements for the K giant $\nu$\,Oph (=~HIP\,88048, HD\,163917, HR\,6698), which reveal two brown dwarf companions with a period ratio close to 6:1. For our orbital analysis we use 150 precise RV measurements taken at Lick Observatory between 2000 and 2011, and we combine them with RV data for this star available in the literature. Using a stellar mass of $M = 2.7$\,$M_\odot$ for $\nu$\,Oph and applying a self-consistent N-body model we estimate the minimum dynamical companion masses to be $m_1\sin i \approx 22.2\,M_{\mathrm{Jup}}$ and $m_2\sin i \approx 24.7\,M_{\mathrm{Jup}}$, with orbital periods $P_1 \approx 530$\,d and $P_2 \approx 3185$\,d.
We study a large set of potential orbital configurations for this system, employing a bootstrap analysis and a systematic $\chi_{\nu}^2$ grid-search coupled with our dynamical fitting model, and we examine their long-term stability. We find that the system is indeed locked in a 6:1 mean motion resonance (MMR), with $\Delta \omega$ and all six resonance angles $\theta_{1}, \ldots, \theta_{6}$ librating around~0$^\circ$. We also test a large set of coplanar inclined configurations, and we find that the system will remain in a stable resonance for most of these configurations. The $\nu$\,Oph system is important for probing planetary formation and evolution scenarios. It seems very likely that the two brown dwarf companions of $\nu$\,Oph formed like planets in a circumstellar disk around the star and have been trapped in a MMR by smooth migration capture.
}

  \titlerunning{Two brown dwarfs in 6:1 mean motion resonance}
  \authorrunning{A. Quirrenbach, T. Trifonov, M.H. Lee, S. Reffert}
   \maketitle
   %\titilerunning%\authorrunning
%________________________________________________________________

\section{Introduction}

The generally accepted concept of planet formation suggests that
orbital mean motion resonances (MMR) among planetary systems are most likely established during the early stages of planet formation.
This is possible since the newly formed proto-planets will undergo
considerable gravitational interactions with the circumstellar disk from which they have formed.
These planet-disk interactions lead to differential planet migration
inward or outward from their birthplaces \citep{Bryden2000, Kley2000, Lee2002} until the disk dissipates and
planets reach their final orbital configurations.
Such a scenario allows planets with mutually widely separated orbits to approach each other until
their orbits are slowly synchronized and trapped into a mean motion resonance with a period ratio close to a ratio of small integers.

During the past two decades of Doppler exoplanet surveys\footnote{
Main sequence: \citet{Mayor1995, Butler1997, Fischer2007} etc.,
 Sub-giants: \citet{Johnson2006} etc.,
 and Giants: \citet{Frink, Setiawan2003, Sato2003, Niedzielski2007, Reffert2015} etc.}
 a significant number of mean motion resonant pair candidates have been found.
The diversity of period ratios in extrasolar multi-planet systems covers many possible configurations;
these systems are found with MMR close to
2:1 \citep{Marcy, Trifonov2014}, 3:2 \citep{Correia}, 3:1 \citep{Desort2008} and even 5:1 \citep{Correia2005}. From these Doppler surveys it was inferred early on that brown dwarfs (i.e., objects with minimum masses between the deuterium burning limit at $\sim 13\,M_{\rm Jup}$ and the hydrogen burning limit at $\sim 70\,M_{\rm Jup}$) are not very abundant as companions to Solar-type stars \citep{Marcy2000, Grether2006, Sahlmann2011}. However, massive planets and brown dwarfs are rather common companions to giant stars, many of which have masses considerably larger than 1\,$M_\odot$ \citep{Sato2012, Mitchell2013, Reffert2015}.

Brown dwarf companions to Solar-type and more massive stars may form through two distinct channels: One potential formation mechanism is gravitational collapse inside the molecular cloud; in this case the star-brown dwarf pair may be regarded a stellar binary with extreme mass ratio. Alternatively, brown dwarfs could be formed in a massive circumstellar disk, in a manner similar to the most massive gas giant planets. (We may of course also observe a mix of objects formed in either way.) We will argue that a brown dwarf system locked in a mean motion resonance presents a strong argument in favor of the notion that brown dwarfs can actually be formed in a proto-planetary disk and thus be regarded as ``Super Jupiter'' planets.

In this paper we introduce our Doppler measurements for the double-brown dwarf system orbiting around the intermediate mass K giant star $\nu$\,Oph.
This star is accompanied by two brown dwarfs with orbital period ratio very close to 6:1.
Here for a first time we introduce a detailed orbital analysis of the $\nu$\,Oph system,
and we study its resonance configuration. To our knowledge this ``super-planet'' resonant system is the first of its kind
and represents an important clue on the scenario of planet and brown dwarf formation within a disk.

\begin{table}[tp]

\caption{Stellar properties for HB and RGB models of $\nu$\,Oph.}
\label{table:phys_param}

\begin{tabular}{ l p{1.3cm} r@{ }l r@{ }l }     % 2 columns

\hline\hline  \noalign{\vskip 0.8mm}
  Parameter   & & \multicolumn{2}{c}{HB} & \multicolumn{2}{c}{RGB}  \\%&  reference \\
\hline    \noalign{\vskip 0.8mm}
   %Spectral type                           & & K0III  &           & K0III  &              \\
   Probability                             & & \multicolumn{2}{c}{>99\%}  &   \multicolumn{2}{c}{<1\%} \\
   Mass    [$M_{\odot}$]                   & & 2.7    & $\pm$ 0.2 & 3.0    & $\pm$ 0.2   \\
   Luminosity    [$L{_\odot}$]             & & 109.3  & $\pm$ 3.1 & 109.0  & $\pm$ 3.9    \\
   Radius    [$R_{\odot}$]                 & & 14.6   & $\pm$ 0.3 & 14.3   & $\pm$ 0.6    \\
   Age    [Gyr]                        & & 0.65   & $\pm$ 0.17 & 0.39   & $\pm$ 0.10   \\
   $T_{\mathrm{eff}}$~[K]                  & & 4886   & $\pm$ 42  & 4936   & $\pm$ 95     \\
   $\log g~[\mathrm{cm\cdot s}^{-2}]$      & & 2.56   & $\pm$ 0.04& 2.63   & $\pm$ 0.07   \\
  % Fe/H [dex]                              & & 0.06   & $\pm$ 0.1 & 0.06   & $\pm$ 0.1    \\
%   RV$_{absolute}$   $[$km\,s$^{-1}]$      & & 12.95  &              \\
\hline\hline
%   $\alpha$ - This paper, $\beta$ - other paper\\

\end{tabular}
%\end{minipage}%}
\end{table}

We structure the paper as follows:
in Sect.~\ref{HIP88048} we give a brief description on what is already known in the literature
for $\nu$\,Oph and its sub-stellar companions and we present our full set of precise
radial velocities taken at Lick Observatory.
In Sect.~\ref{Orbital fit} we describe our data analysis strategy and we introduce our
coplanar and inclined N-body dynamical models to
the available Doppler data from Lick and from \citet{Sato2012}.
%and we discuss the dynamical properties of these fits.
We discuss our best-fit parameter error estimation techniques in Sect.~\ref{Error}, while
in Sect.~\ref{Stability} we study the $\nu$\,Oph system's dynamical and statistical properties
in the orbital phase space around the best fit based on a systematic $\chi_{\nu}^2$ grid-search
analysis. In Sect.~\ref{Discussion} we discuss the implications of our findings in the context of formation scenarios for planets and brown dwarfs. Finally, in Sect.~\ref{Summary} we summarize our results.

%
%
% \begin{table}[!htp]
%
%
%
% \caption{Stellar properties of $\nu$\,Oph }
% \label{table:phys_param}
%
% \begin{tabular}{ l p{2.7cm} r@{ }l }     % 2 columns
%
% \hline\hline  \noalign{\vskip 0.8mm}
%   Parameter   & & $\nu$\,Oph  \\%&  reference \\
% \hline    \noalign{\vskip 0.8mm}
%    Spectral type                           & & K0III  &              \\
%    Age    $[$Gyr$]$                        & & 0.65   & $\pm$ 0.2   \\
%    Mass    [$M_{\odot}$]                   & & 2.7    & $\pm$ 0.2   \\
%    Luminosity    [$L{_\odot}$]             & & 109.3  & $\pm$ 3.1    \\
%    Radius    [$R_{\odot}$]                 & & 14.6   & $\pm$ 0.3    \\
%    $T_{\mathrm{eff}}$~[K]                   & & 4886   & $\pm$ 42     \\
%    $\log g~[\mathrm{cm\cdot s}^{-2}]$      & & 2.56   & $\pm$ 0.04   \\
%    Fe/H [dex]                                & & 0.06   & $\pm$ 0.1    \\
% %   RV$_{absolute}$   $[$km\,s$^{-1}]$      & & 12.95  &              \\
% \hline\hline
% %   $\alpha$ - This paper, $\beta$ - other paper\\
%
% \end{tabular}
% %\end{minipage}%}
% \end{table}
%

\section{$\nu$\,Oph and its companions}
\label{HIP88048}

\subsection{Stellar parameters}
\label{Stellar parameters}

$\nu$\,Oph (=~HIP\,88048, HD\,163917, HR\,6698) is a bright ($V = 3.32$\,mag) photometrically stable
K0III giant star (variability $\leq3$\,mmag) at a distance of $46.2 \pm 0.6$\,pc \citep{Leeuwen}.
Stellar parameters for this star were estimated following \citet{Reffert2015} and \citet{Stock2018}:
Knowing from Hipparcos data\footnote{For the very bright star $\nu$\,Oph, the Hipparcos parallax is more precise than that from Gaia DR2.} the position of $\nu$\,Oph in the Hertzsprung-Russell (HR) diagram,
we constructed theoretical evolutionary tracks and isochrones \citep{Girardi2000}.
However, since the positions of the evolutionary tracks and stellar isochrones
in the HR diagram also depend on the primordial stellar chemical abundance we include the stellar metallicity
as an additional parameter in the trilinear model interpolation.
Considering $\nu$\,Oph's measured color ($B - V = 0.987 \pm 0.035$), absolute magnitude ($M_V = - 0.19 \pm 0.04$)
and metallicity \citep[{$\mathrm{[Fe/H]} = 0.06 \pm 0.1$;}][]{Hekker2} we generated 10\,000 positions in the HR diagram
consistent with the obtained uncertainties on these quantities and for each position
we estimated effective temperature $T_{\mathrm{eff}}$, stellar mass $M$, luminosity $L$, radius $R$, and
surface gravity $\log g$.
From these estimates we determined the most probable stellar parameters and their uncertainties,
along with the probability of $\nu$\,Oph being on the red giant branch (RGB) or the horizontal branch (HB), respectively.

%star with photometric variation lower than the 3~mmag.
%\citet{Hekker2} estimated that $\nu$\,Oph has near solar metallicity of [Fe/H] = 0.06 $\pm$ 0.1,
%Based on the stellar metallicity estimated in \citet{Hekker2} ([Fe/H] = 0.06 $\pm$ 0.1)
%Given the estimated star metallicity ([Fe/H] = 0.06 $\pm$ 0.1) and it's
%observed position on the HR diagram, \citet{Reffert2014}

We find that $\nu$\,Oph is most likely an intermediate-mass horizontal branch star with an
age of $0.65 \pm 0.17$\,Gyr, stellar mass $M = 2.7 \pm 0.2\,M_{\odot}$, luminosity of $L = 109.3 \pm 3.1\,L_{\odot}$, radius of
$R = 14.62 \pm 0.32\,R_{\odot}$, and an effective temperature of $T_{\mathrm{eff}} = 4886 \pm 42$\,K.
%Yet, we find that there is 19\% chance $\nu$\,Oph to be located at the RGB
%being younger (0.39~$\pm$~0.10 Gyr) and slightly more massive with stellar mass of $M$~=~3.02~$\pm$~0.18~$M_{\odot}$.
%$Here we adopt the stellar parameters with the highest probability, namely the HB estimates.
Stellar parameters for HB and RGB models of $\nu$\,Oph are summarized in Tab.~\ref{table:phys_param},
while additional physical parameter estimates for this star are given in \citet{Reffert2015} and \citet{Stock2018}.

  \begin{figure}[tp]

    \includegraphics[width=9cm]{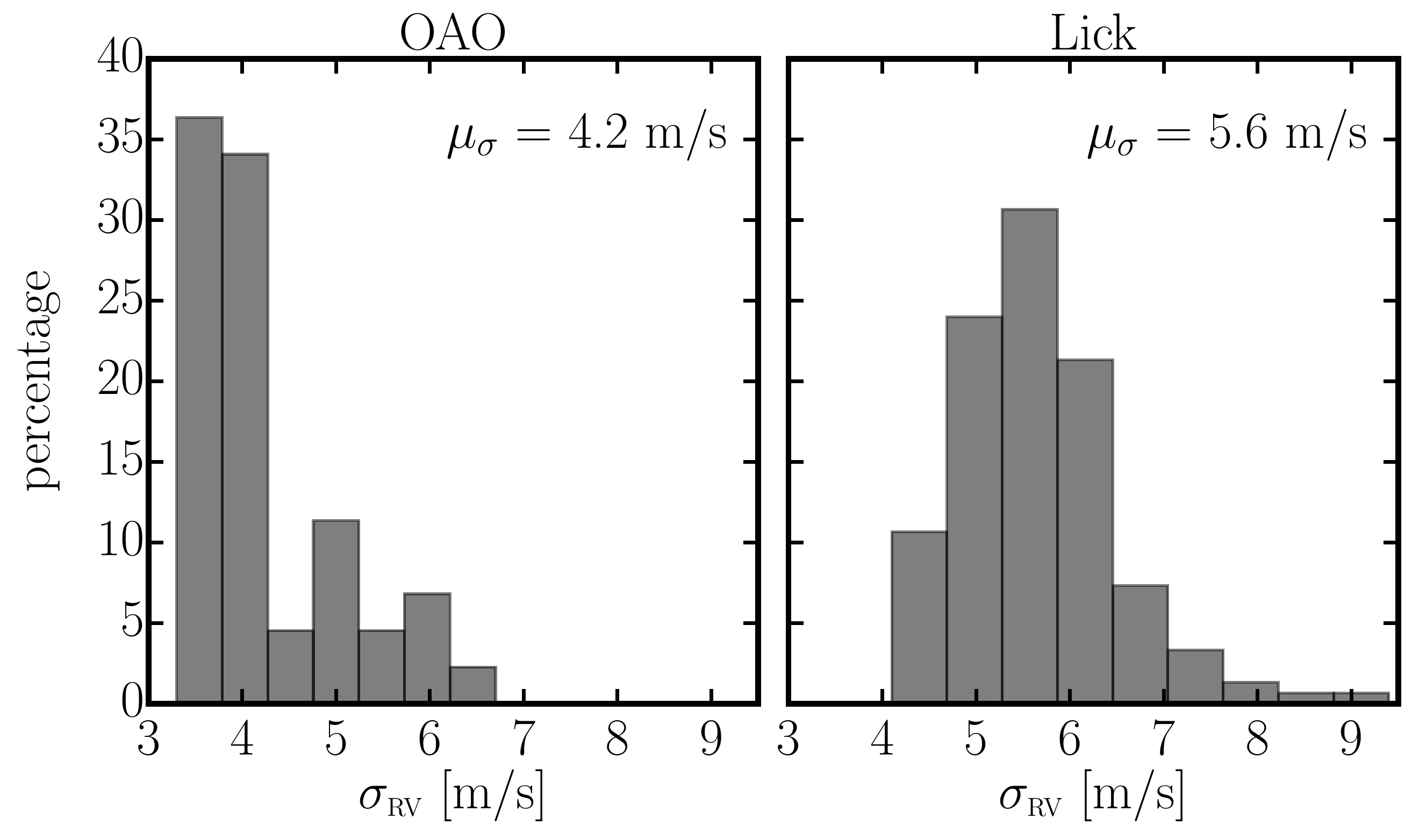}

  \caption{Distribution of the uncertainties for the OAO and Lick radial velocities.
  The OAO data set has only 44 RVs, but they are slightly more precise than the 150 RVs from Lick.
}
  \label{data_stat}
  \end{figure}

\subsection{Radial velocity data}
\label{RVdata}

We extensively monitored $\nu$\,Oph for Doppler variations at UCO
Lick  Observatory between November 2000 and November 2011.
We collected a total of 150 precise stellar radial velocity measurements
using the Iodine cell method \citep{Valenti, Butler} in conjunction with the
Hamilton spectrograph \citep[R~$\approx$~60\,000:][]{Vogt1987} and the 0.6\,m Coud\'{e} Auxiliary Telescope (CAT).
Since $\nu$\,Oph is a bright star the typical exposure times with the CAT were about 300 seconds,
which results in a signal-to-noise ratio of about S/N $\sim$ 120$-$140.
Our precise velocities are obtained in the wavelength region between 5000 and 5800\,\AA{}
where most of the calibration iodine lines are superimposed onto the stellar spectra.
For $\nu$\,Oph we achieved a typical velocity precision of 4 to 9\,m\,s$^{-1}$,
with an estimated mean precision of 5.6\,m\,s$^{-1}$ (see Fig.~\ref{data_stat}).
All radial velocities (RVs) from Lick and their estimated formal errors are given in Tab.~\ref{table:rvlick}.

The observations of $\nu$\,Oph were taken
as part of our Lick Doppler survey of 373 very bright ($V \leq 6$\,mag) G and K giants \citep{Frink}.
%For these stars we obtained precise (5 -- 8~m\,s$^{-1}$) radial velocities using the Hamilton spectrograph
% (R $\approx$ 60\,000) in conjunction with the Iodine cell method \citep{Valenti,Butler}.
Our primary program objective is to investigate the planet occurrence and evolution around intermediate-mass stars as
function of stellar mass, metallicity and evolutionary stage of the stars.
Our targets were selected from the Hipparcos Catalogue \citep{ESA1997} with the condition to be
photometrically constant single stars with estimated
stellar masses between 1 and 5\,$M_{\odot}$.
The program and star selection criteria are described in more details in \citet{Frink},
while planetary companions from our survey were published in \citet{Frink2}, \citet{Reffert},
\citet{Quirrenbach}, \citet{Mitchell2013}, \citet{Trifonov2014}, and \citet{Ortiz2016}.
Results from the planet occurrence rate from our G and K giant sample
have been published in \citet{Reffert2015}.

 \begin{figure*}[tp!]
 \begin{center}$
\begin{array}{ccc}

    \includegraphics  [width=11cm]{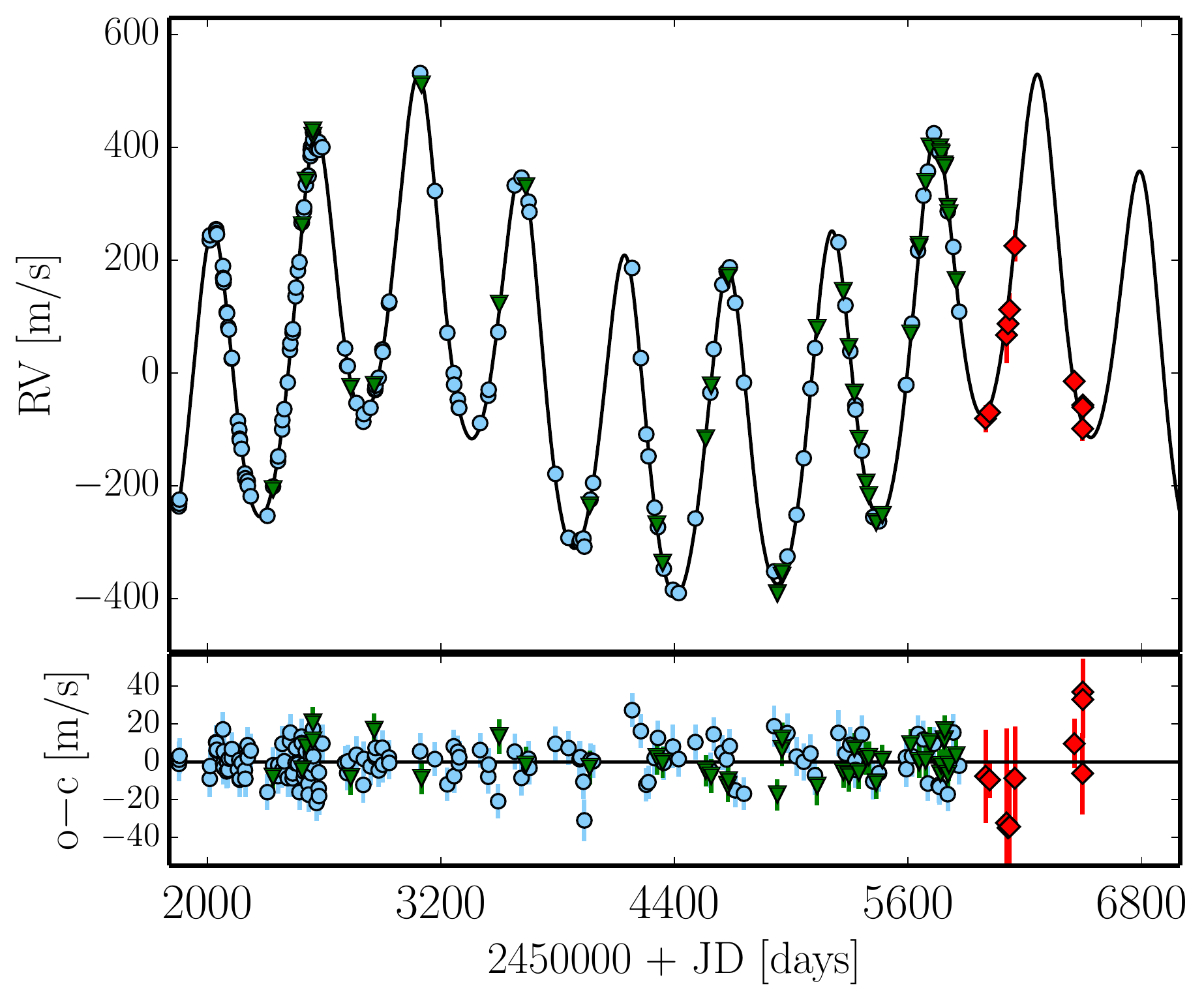}\\

    \end{array}$
 \end{center}

\caption{Radial velocities for $\nu$\,Oph along with their
error bars measured at Lick Observatory are shown with blue circles covering more than 11 years from July 2000 to October 2011.
Green triangles denote the velocities from OAO, while the VLT near-IR data are plotted with red diamonds.
The solid line illustrates the best dynamical model applied to the combined optical data from Lick and OAO.
The near-IR data from VLT is only superimposed on the dynamical model with the only fitting parameter being an RV offset for the whole data set.
%and the best-fit is overplotted respect to the optical fit.
The two wavelength domains are clearly consistent with each other. The bottom panel shows the residuals around the best fit.}
  \label{rv_plot}
  \end{figure*}

Based on the initial RV data from our survey \citet{Mitchell2003}
reported the discovery of a brown dwarf companion around $\nu$\,Oph
with $m_1 \sin i \approx 22.2\,M_{\mathrm{Jup}}$ and an orbital period of $P_1 \approx 530$\,d.
Follow-up observations at Lick showed that a single Keplerian fit cannot explain the data well.
%By the time we finished our observations at Lick the full RV data set clearly
By mid-2010 the Lick RV data set clearly revealed the presence of a second longer-period
sub-stellar companion with a minimum mass consistent with a brown dwarf object.
Based on 135 Lick RVs, the two brown dwarf system was announced in \citet{Quirrenbach}, who gave orbital parameters of
$P_1 \approx 530$\,d, $m_1 \sin i \approx 22.3\,M_{\mathrm{Jup}}$, $e_1 \approx 0.13$ and
$P_2 \approx 3170$\,d, $m_2 \sin i \approx 24.5\,M_{\mathrm{Jup}}$, $e_2 \approx 0.18$, respectively.
(In this paper we use the label ``1'' for the inner companion $\nu$\,Oph\,b, and ``2'' for the outer companion $\nu$\,Oph\,c.)
To our knowledge \citet{Quirrenbach} was the first reported case of a star orbited by two brown dwarfs.
Even more remarkably, the reported period ratio between the two orbits appears to be very close to 6:1,
which suggests that the system might be locked in a mean motion resonance.

A year later the $\nu$\,Oph system was confirmed by \citet{Sato2012}
based on 44 precise RVs from the Okayama Astrophysical Observatory (OAO), Japan, collected
between February 2002 and July 2011.
Similar to our Lick program, the OAO observations were carried out
with an iodine absorption cell mounted on the
HIDES Spectrograph \citep[R~$\approx$~67\,000:][]{Izumiura1999} at a larger aperture 1.88\,m telescope.
The OAO data have slightly better precision than our Lick data with mean precision of 4.2\,m\,s$^{-1}$;
most likely as a result of a higher signal-to-noise ratio reached for the HIDES spectra
\citep[S/N $\sim$ 200, see][]{Sato2012}.
A comparison of the formal precision of the data sets from Lick and OAO is shown in Fig.~\ref{data_stat}.
The Keplerian spectroscopic orbital parameters for the $\nu$\,Oph system reported in
 \citet{Sato2012} are in general agreement with those from \citet{Quirrenbach}.
However, \citet{Sato2012} adopted a larger stellar mass for $\nu$\,Oph equal to $M = 3.04\,M_{\odot}$,
and thus they derived slightly higher masses for the companions, with minimum masses of
$m_1 \sin i \approx 24.0\,M_{\mathrm{Jup}}$ and $m_2 \sin i \approx\,27.0\,M_{\mathrm{Jup}}$, respectively.

In addition to the velocities from Lick and OAO obtained at optical wavelengths,
we collected a total of 10 near-IR absolute RVs with the CRIRES
spectrograph at the VLT between 2011 and 2013 \citep{Trifonov2015}.
The aim of this test was to confirm or disprove the planetary origin
of the radial velocity signals for 20 of our Lick stars, including $\nu$\,Oph.
%The RVs from CRIRES are given in Trifonov et al. (2015),

The CRIRES data cannot be used to further constrain
the orbital configuration. This is mostly because of the very limited phase coverage of the CRIRES data
compared to the OAO and Lick data sets, the lack of temporal overlap between the data sets in two wavelength domains,
and the relatively low near-IR velocity precision ($\sim 25$\,m\,s$^{-1}$), when compared with the optical data.
However, despite the incomplete phase coverage, using the near-IR data alone we were able to construct
one full period of the inner companion (see Fig.~\ref{rv_plot}).
We find that the near-IR data are fully consistent with the best-fit prediction based on the optical data,
and therefore there can be little doubt on the companion hypothesis for $\nu$\,Oph.

\subsection{Stellar jitter}
\label{jitter}

From our full Lick RV data set, we estimate the additional
astrophysical RV noise around the best fit to be $\sim 7.5$\,m\,s$^{-1}$.
In this paper we adopt this short-term velocity scatter as stellar ``jitter'' \citep{Wright2005, Hekker}.
In fact, the same jitter level was estimated by \citet{Sato2012} using their OAO data set.
This stellar jitter amplitude for $\nu$\,Oph is typical for other late G and early K giants
and is most likely due to rapid solar-like $p$-mode oscillations \citep{De_Ridder, Barban, Zechmeister2008},
which appear as RV noise in our data.
Based on the physical properties of $\nu$\,Oph and the scaling relation from \citet{Kjeldsen1995},
we estimate a scatter velocity of $\sim 9$\,m\,s$^{-1}$ and period of 0.27\,d,
which agrees well with our result from the RV data.
In a more complete analysis we could have left the jitter as a free parameter and estimated it together with the other system parameters, but as we are not interested in the exact value and an error estimate for the jitter term, we keep it fixed for simplicity. This is not expected to have a significant influence on any of the other results, as a reduced $\chi_\nu^2 \approx 1$ would be the preferred outcome in any case.

%\section{Orbital fitting and stability setup}
\section{Best fits}

\label{Orbital fit}

To model the orbital configuration of $\nu$\,Oph we adopted a stellar mass of $M = 2.7\,M_\odot$
and we used all the available optical RV data for $\nu$\,Oph.
We combined our Lick radial velocities with those published in \citet{Sato2012},
resulting in a total of 194 precise RVs with typical uncertainties of the order of 3-9\,m\,s$^{-1}$.
For both data sets we quadratically added the estimated RV jitter of 7.5\,m\,s$^{-1}$
into the total error budget, and hence we considered the astrophysical stellar noise
as an additional RV uncertainty.
We analyzed the combined RV data by adopting the methodology described in \citet{Tan2013}.
We applied a Levenberg-Marquardt (LM) based $\chi^2$ minimization technique
coupled with two models, namely a double-Keplerian model and a
self-consistent dynamical model with the equations of motion integrated using the Gragg-Bulirsch-Stoer
integration method \citep{Press}.
For both two-planet models the fitted parameters are the spectroscopic elements: radial velocity semi-amplitude~$K$,
orbital period~$P$, eccentricity~$e$, argument of periastron $\omega$,
mean anomaly~$M_0$ and the RV data offset RV$_{\mathrm{off}}$ for each data set.
All orbital parameters (including the derived semi-major axes $a_1$, $a_2$ and
minimum masses $m_1 \sin i$, $m_2 \sin i$) are obtained in the Jacobi frame \citep[e.g.][]{Lee2003} and are
valid for the first observational epoch, which in our case is always JD\,2451853.595
(the first Lick data point).

We further test our models for long-term dynamical stability
using the {\it SyMBA} symplectic integrator \citep{Duncan1998}.
The {\it SyMBA} integrator was modified to work directly
with the obtained Jacobi elements as an input and we were able to
simultaneously monitor the evolution of the orbital elements over time.
Since we were aware that the companions period ratio is close to 6:1 we
additionally monitored the evolution of the secular apsidal angle
$\Delta\omega = \omega_1 - \omega_2$
and the evolution of all six resonance angles, defined as:

 \begin{flalign}
  &  \theta_1 =  \lambda_1 - 6\lambda_2 + 5\varpi_1, & \\
  &  \theta_2 =  \lambda_1 - 6\lambda_2 + 4\varpi_1 + \varpi_2,   &  \\
  &  \theta_3 =  \lambda_1 - 6\lambda_2 + 3\varpi_1 + 2\varpi_2,  &   \\
  &  \theta_4 =  \lambda_1 - 6\lambda_2 + 2\varpi_1 + 3\varpi_2,  & \\
  &  \theta_5 =  \lambda_1 - 6\lambda_2 + \varpi_1 + 4\varpi_2,   & \\
  &  \theta_6 =  \lambda_1 - 6\lambda_2 + 5\varpi_2  &
\end{flalign}

%\begin{equation}
%\theta_{m=1,6} =  6\lambda_2  - 1\lambda_2 - (6-m)\varpi_1  +
%(1-m)\varpi_2 = 6M_2 - M_1 + (6 - m + 1)\Delta\varpi,
%\label{eq:theta}
%\end{equation}

where $\varpi_{1,2} = \Omega_{1,2} + \omega_{1,2}$ is the longitude of
periastron and
$\lambda_{1,2} = M_{1,2} + \varpi_{1,2}$ is the mean longitude of
the inner and outer companion, respectively.
Clearly, $\theta_{\rm 1 \ldots 6}$ are not independent and can be derived
from the evolution of one of the resonance angles and $\Delta\varpi$.
The expansion of all possible resonance angles, however, helps to
identify the resonance angle $\theta_n$ with the lowest libration
amplitude, which is an important dynamical characteristic of the system.

 \begin{figure*}[tp!]
 \begin{center}$
\begin{array}{ccc}
    \includegraphics  [width=18cm]{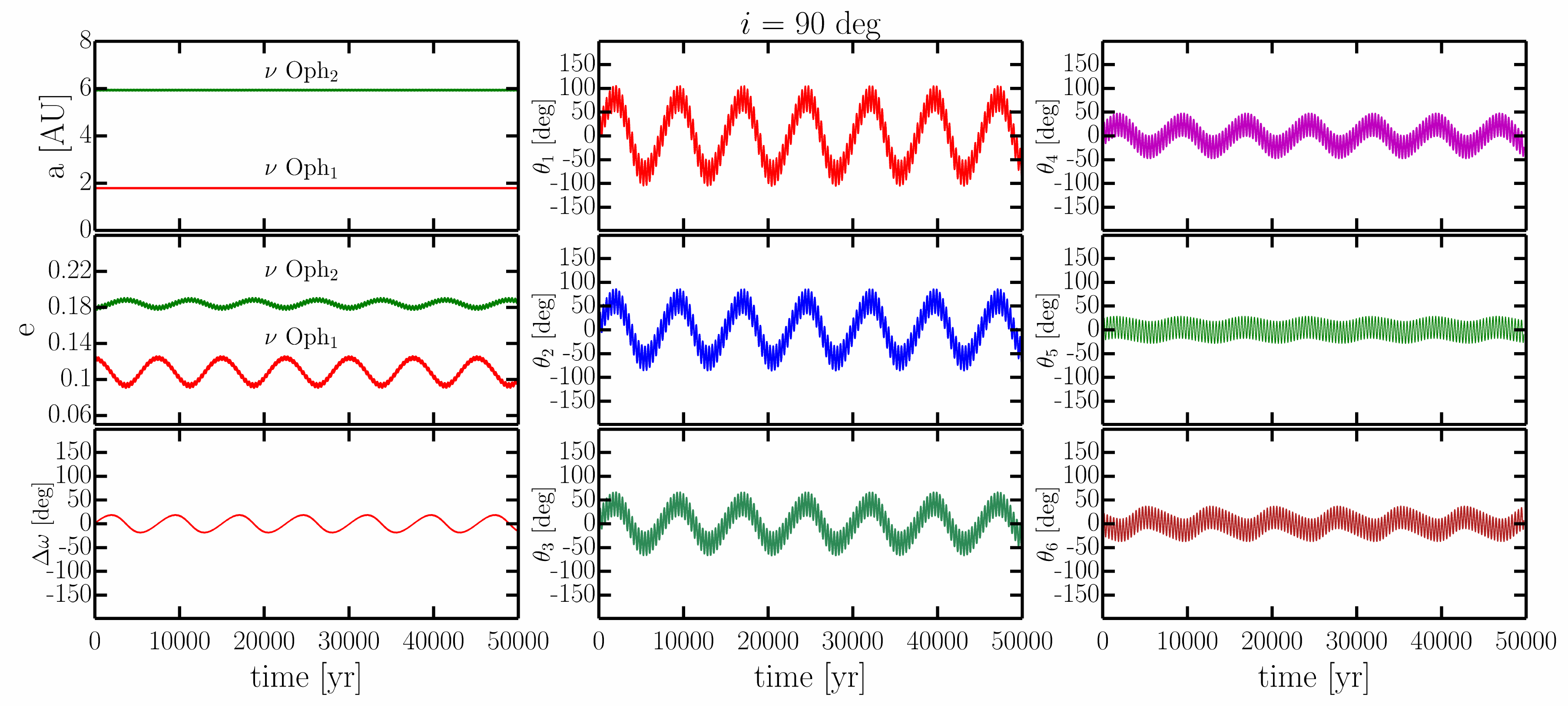}\\
    \end{array}$
 \end{center}

\caption{Best coplanar edge-on orbital evolution shown for 50\,kyrs from the numerical simulation.
{\it Left:} From top to bottom are shown the orbital evolution of the companions' semimajor
axes, eccentricities and the secular apsidal angle $\Delta\omega$. Clearly no change in $a_1$ and $a_2$
can be seen, while $e_1$ and $e_2$ are librating in opposite phase with small amplitudes.
The other panels show the resonance angles $\theta_{1,2,3}$ and $\theta_{4,5,6}$, which are
clearly librating around 0$^\circ$ indicating the resonant nature of the system. See text for more details.
%$Middle:$ largest amplitude has $\theta_1 = 105.3^\circ$, followed by  $\theta_2 = 86.0^\circ$ and
%$\theta_3 = 66.8^\circ$. $Right:$ is $\theta_4 = 47.6^\circ$, $\theta_5 = 28.9^\circ$ and
%$\theta_6 = 37.3^\circ$.
}
  \label{theta_evol}
  \end{figure*}

\begin{table}[!ht]

%\resizebox{0.75\textheight}{!}
%{\begin{minipage}{1.1\textwidth}

\centering
\caption{Best coplanar fits for the combined data sets of $\nu$\,Oph.}
\label{table:orb_par_stable}

\begin{tabular}{ lccccc}     % 2 columns

\hline\hline  \noalign{\vskip 0.7mm}

      \multicolumn{3}{c}{ Keplerian}            \\

\hline \noalign{\vskip 0.7mm}

Orb. Param.& $\nu$\,Oph\,b & $\nu$\,Oph\,c  \\

 \hline\noalign{\vskip 0.5mm}

   $K$  [m\,s$^{-1}$]                        & 288.27 $\pm$ 0.98     &  ~~176.78  $\pm$ 1.30      \\
   $P$ [d]   			     & 530.02 $\pm$ 0.11     &   3183.02 $\pm$ 5.89       \\
   $e$                                       & ~~~~0.124 $\pm$ 0.003 &  ~~~~~~0.180 $\pm$ 0.006   \\
   $\omega$ [$^\circ$]                            & ~~~~9.88 $\pm$ 1.50   &  ~~~~~~8.57 $\pm$ 1.97     \\
   $M_0$ [$^\circ$]                                 & 235.78 $\pm$ 1.52     &  ~~222.75 $\pm$ 1.92       \\  \noalign{\vskip 0.9mm}
   RV$_{ \mathrm{off.~Lick}}$~[m\,s$^{-1}$]   &\multicolumn{2}{c}{$-$49.32 $\pm$ 1.83}                                \\ \noalign{\vskip 0.9mm}
   RV$_{ \mathrm{off.~OAO}}$~[m\,s$^{-1}$]    &\multicolumn{2}{c}{~~~~0.34 $\pm$ 1.74}                               \\ \noalign{\vskip 0.9mm}
%   $i$ [deg]                                 && 90.0                  &                          && 16.0                  &          \\
%   $\Delta\Omega$ [deg]                      && 0.0                   &                          && 0.0                   &           \\ \noalign{\vskip 0.9mm}
   $a$ [AU]                                  & 1.789                 &   5.929                    \\
   $m \sin i$ [$M_{\mathrm{Jup}}$]                  & 22.204                &   24.674                   \\ \noalign{\vskip 0.9mm}
 %  jitter [m\,s$^{-1}$]                      & 7.5                   &                            \\
   $r.m.s. $ [m\,s$^{-1}$]                   & \multicolumn{2}{c}{9.21}                                             \\
   $\chi_{\nu}^2$                            & \multicolumn{2}{c}{1.0456}                                            \\
   \noalign{\vskip 0.5mm}

 \hline\hline\noalign{\vskip 1.2mm}

   \multicolumn{3}{c}{N-body ($i$~=~90$^\circ$, $\Delta\Omega$~=~0$^\circ$) } \\

\hline \noalign{\vskip 0.7mm}

Orb. Param.& $\nu$\,Oph\,b & $\nu$\,Oph\,c  \\% &$\nu$\,Oph$_{\mathrm{1}}$ &$\nu$\,Oph$_{\mathrm{2}}$ \\  %\noalign{\vskip 0.4mm}
 \hline\noalign{\vskip 0.5mm}
   $K$  [m\,s$^{-1}$]                        & 288.26 $\pm$ 0.99     &  ~~176.73  $\pm$ 1.24      \\
   $P$ [d]   			     & 530.21 $\pm$ 0.10     &   3184.83 $\pm$ 5.93       \\
   $e$                                       & ~~~~0.124 $\pm$ 0.003 &  ~~~~~~0.180 $\pm$ 0.006   \\
   $\omega$ [$^\circ$]                            & ~~~~9.93 $\pm$ 1.49   &  ~~~~~~8.27 $\pm$ 1.98     \\
   $M_0$ [$^\circ$]                                 & 235.69 $\pm$ 1.52     &  ~~223.02 $\pm$ 1.92       \\  \noalign{\vskip 0.9mm}
   RV$_{ \mathrm{off.~Lick}}$~[m\,s$^{-1}$]   &\multicolumn{2}{c}{$-$48.63 $\pm$ 0.95}                                \\ \noalign{\vskip 0.9mm}
   RV$_{ \mathrm{off.~OAO}}$~[m\,s$^{-1}$]    &\multicolumn{2}{c}{~~~~0.33 $\pm$ 1.74}                               \\ \noalign{\vskip 0.9mm}
%   $i$ [deg]                                 && 90.0                  &                          && 16.0                  &          \\
%   $\Delta\Omega$ [deg]                      && 0.0                   &                          && 0.0                   &           \\ \noalign{\vskip 0.9mm}
   $a$ [AU]                                  & 1.790                 &   5.931                    \\
   $m$ [$M_{\mathrm{Jup}}$]                  & 22.206                &   24.662                   \\ \noalign{\vskip 0.9mm}
 %  jitter [m\,s$^{-1}$]                      & 7.5                   &                            \\
   $r.m.s. $ [m\,s$^{-1}$]                   & \multicolumn{2}{c}{9.18}                                              \\
   $\chi_{\nu}^2$                            & \multicolumn{2}{c}{1.0414}                                            \\
   \noalign{\vskip 0.5mm}

 \hline\hline\noalign{\vskip 1.2mm}

   \multicolumn{3}{c}{N-body ($i$~=~16$^\circ$, $\Delta\Omega$~=~0$^\circ$) } \\

\hline \noalign{\vskip 0.7mm}

Orb. Param.& $\nu$\,Oph\,b & $\nu$\,Oph\,c  \\% &$\nu$\,Oph$_{\mathrm{1}}$ &$\nu$\,Oph$_{\mathrm{2}}$ \\  %\noalign{\vskip 0.4mm}
 \hline\noalign{\vskip 0.5mm}
   $K$  [m\,s$^{-1}$]                        & 288.21 $\pm$ 0.97     &  175.23  $\pm$ 1.20        \\
   $P$ [d]   			     & 530.73 $\pm$ 0.10     &  3188.95 $\pm$ 6.26        \\
   $e$                                       & ~~~~0.124 $\pm$ 0.003 &  ~~~~~~0.178 $\pm$ 0.006   \\
   $\omega$ [$^\circ$]                            & ~~~~7.59 $\pm$ 1.99   &  ~~~~~~9.74 $\pm$ 2.13     \\
   $M_0$ [$^\circ$]                                 & 235.56 $\pm$ 1.50     & ~~223.64 $\pm$ 1.92        \\  \noalign{\vskip 0.9mm}
   RV$_{ \mathrm{off.~Lick}}$~[m\,s$^{-1}$]   &\multicolumn{2}{c}{$-$49.26 $\pm$ 0.92}                                \\ \noalign{\vskip 0.9mm}
   RV$_{ \mathrm{off.~OAO}}$~[m\,s$^{-1}$]    &\multicolumn{2}{c}{~~~~0.30 $\pm$ 1.74}                               \\ \noalign{\vskip 0.9mm}
%   $i$ [deg]                                 && 90.0                  &                          && 16.0                  &          \\
%   $\Delta\Omega$ [deg]                      && 0.0                   &                          && 0.0                   &           \\ \noalign{\vskip 0.9mm}
   $a$ [AU]                                  & 1.803                 &   6.022                    \\
   $m$ [$M_{\mathrm{Jup}}$]                  & 81.691                &   91.977                   \\ \noalign{\vskip 0.9mm}
 %  jitter [m\,s$^{-1}$]                      & 7.5                   &                            \\
   $r.m.s. $ [m\,s$^{-1}$]                   & \multicolumn{2}{c}{9.15}                                              \\
   $\chi_{\nu}^2$                            & \multicolumn{2}{c}{1.0367}                                            \\
   \noalign{\vskip 0.5mm}

 \hline\hline

\end{tabular}
%\end{minipage}}
\end{table}

%\noindent
%where  $\varpi_{1,2}$~=~$\Omega_{1,2} + \omega_{1,2}$ is the argument of periastron and
%$\lambda_{1,2}$~=~$M_{1,2}$ + $\varpi_{1,2}$ is the mean longitude of the inner and outer planet, respectively.

We start each orbital integration from JD = 2451853.595,
and we collected orbital output from the simulations for every 10 years of integration.
The simulations were interrupted only in case of mutual collisions between the companions
and the star, or if one of the companions is ejected from the system.
We defined an ejection if one of the companions' semimajor axes exceeds 10\,AU
during the integration time, and we defined a collision with the star if
the semimajor axis of one of the planets goes down to 0.1\,AU.
Since the age of the system is estimated to be $\sim 0.65$\,Gyr, % \citep{Reffert2014},
we integrate the individual best dynamical fits for a maximum of 1\,Gyr, which gave us
more than enough time to study the system's long-term stability.
For our stability test we adopted a time step equal to 2 days leading to
about $\sim 260$ steps per complete orbit of the inner companion.
We find that the selected time step was adequate to assure the precise simulation
of the $\nu$\,Oph system.

\subsection{Coplanar edge-on fit}
\label{coplanar}

The best Keplerian fit to the full optical data set has
%$\chi_{\nu}^2$~=~1.046\footnote{We use $\chi_{\nu}^2$~=~$\frac{\chi^2}{DOF}$ , where $DOF$~=~N$_{\mathrm{data}}$
$\chi_{\nu}^2 = 1.046$ and is
clearly consistent with two massive companions in the brown dwarf regime
with $m_1\sin i = 22.2\,M_{\mathrm{Jup}}$ and $m_2\sin i = 24.7\,M_{\mathrm{Jup}}$.
The orbits are non-circular with $e_1 = 0.124 \pm 0.003$ and $e_2 = 0.180 \pm 0.006$, respectively.
The inner companion has an orbital period of $P_1 = 530.0 \pm 0.1$\,d, while the outer companion has
$P_2 = 3183.0 \pm 5.9$\,d, consistent with the earlier findings by
\citet{Quirrenbach} and \citet{Sato2012}. Orbital parameters and uncertainties estimated from the covariance matrix for our best Keplerian fit are given in
Tab.~\ref{table:orb_par_stable}.

We use the best Keplerian model as a good initial guess for our dynamical fitting.
The available RV data cover a bit more than one full period of the outer companion, and thus
any mutual perturbations between the companions should be barely noticeable in the data.
Indeed, despite both companions having large masses in the brown dwarf regime, they are
too far separated in space to mutually influence their orbits strongly on short time scales.
We thus find the difference between a double Keplerian and
self-consistent two planet edge-on dynamical model to be minor and insignificant.
As can be seen from Tab.~\ref{table:orb_par_stable} both edge-on models mutually agree
within the estimated errors, and therefore
we conclude that there is little advantage to use an N-body model over the simpler double Keplerian.
However, since our goal in this paper is to explore the long-term dynamical orbital evolution of the $\nu$\,Oph system for
coplanar edge-on and inclined configurations, we present results based on the dynamical model.

The best coplanar edge-on dynamical fit to the combined Doppler data has $\chi_{\nu}^2 = 1.042$ and
leads to orbital elements of $P_1 = 530.21 \pm 0.10$\,d,
$P_2 = 3184.83 \pm 5.93$\,d, $e_1 = 0.124 \pm 0.003$ and $e_2 = 0.180 \pm 0.006$,
and estimated masses of $m_1 = 22.2\,M_{\mathrm{Jup}}$
and $m_2 = 24.7\,M_{\mathrm{Jup}}$.
This fit also suggests that the $\nu$\,Oph system is in an aligned
orbital configuration with well constrained arguments of
periastron $\omega_1 = 9.9^\circ \pm 1.5^\circ$ and
$\omega_2 = 8.3^\circ \pm 2.0^\circ$.
The other orbital elements and their estimated parameter errors for this fit are
listed in Tab.~\ref{table:orb_par_stable}.

An illustration of the best coplanar edge-on dynamical fit to the $\nu$ Oph
RV data sets is given in Fig.~\ref{rv_plot}.
The blue circles show the 150 radial velocity data points from Lick,
while the 44 RVs from OAO are shown as green~triangles.
Both data sets cover more than 11 years of observations, which
is slightly more than 1.25 orbital periods of the outer companion.
In addition, the near-IR RVs from CRIRES taken by \citet{Trifonov2015} at later epochs are shown in Fig.~\ref{rv_plot} with red diamonds.
The near-IR data points were superimposed on the best fit from the visible-light data, fitting only an RV offset for the whole near-IR data set.
We use these data only to demonstrate the consistency between the two wavelength domains,
but we did not use them in the orbital analysis.
%This is mostly because the near-IR data have small phase coverage, to
%only about one period of the inner companion. Besides that the lack of overlap
%between the optical and near-IR data samples and the relatively large near-IR uncertainties
%makes the CRIRES data contribution to the best-fit negligible.

The long-term dynamical simulation of the edge-on N-body fit shows that the system is stable and is
indeed locked in a 6:1 MMR with all six resonant angles $\theta_1, \ldots, \theta_6$ librating around 0$^\circ$.
Fig.~\ref{theta_evol} shows a 50\,kyr zoom from the 1\,Gyr dynamical evolution of the best coplanar edge-on dynamical fit.
The left panel from top to bottom illustrates the evolution of the brown dwarfs' semimajor axes ($a_1$, $a_2$),
eccentricities ($e_1$, $e_2$), and the secular apsidal angle $\Delta\omega = \omega_1 - \omega_2$.
Clearly, the semimajor axes do not exhibit any notable variations for 50\,kyr and we find that
this is also the case for the complete orbital simulation time of 1\,Gyr.
During the orbital evolution the semimajor axes oscillate with very low and regular amplitude
around $a_1 \approx 1.8$\,AU and $a_2 \approx 5.9$\,AU,
while the orbital eccentricities oscillate in off-phase fashion.
The inner brown dwarf has a mean orbital eccentricity of $e_1 \approx  0.11$, varying between 0.09 and 0.13, while
the outer exhibits a lower eccentricity amplitude between 0.18 and 0.19 with a mean of $e_2 \approx 0.185$.
The secular resonance angle $\Delta \omega$ clearly librates around 0$^\circ$ with semiamplitude of about
20$^\circ$, showing that the system remains in an aligned configuration during the dynamical test.
The middle and the right panels of Fig.~\ref{theta_evol} show the
evolution of the resonant angles $\theta_1$, $\theta_2$, $\theta_3$ (from top to bottom)
and $\theta_4$, $\theta_5$, $\theta_6$, respectively.
For the best dynamical fit the largest semi-amplitude has $\Delta \theta_1 = 105.3^\circ$,
followed by $\Delta \theta_2 = 86.0^\circ$, $\Delta \theta_3 = 66.8^\circ$, $\Delta\theta_4 = 47.6^\circ$, and
the smallest libration semi-amplitude is at $\Delta \theta_5 = 28.9^\circ$. The last resonant angle $\Delta\theta_6$
librates with a semi-amplitude of 37.3$^\circ$, out of phase with respect to $\theta_{1}, \ldots, \theta_{5}$.
The resonant libration period of the eccentricities and all resonance angles is about $\sim 7$\,kyr,
while these also exhibit lower amplitude and very regular short-period variations.

We note that the pattern depicted in Fig.~\ref{theta_evol}, with $\theta_5$ having the smallest libration amplitude of all resonance angles, and a phase reversal between $\theta_5$ and $\theta_6$, is observed robustly over the parameter space that we have investigated. This behavior therefore provides a constraint on the resonance capture mechanism, which must be reproduced by models of the early evolution of the system.

 \begin{figure}[tp!]
 \begin{center}%$
%\begin{array}{ccc}
 % \includegraphics  [width=9cm]{\pathh  hip88048/incl_test_comb/HIP88048_incl_reduced_chi_both.pdf}\\
   %   \includegraphics  [width=9cm]{\pathh  hip88048/incl_test_sato/HIP88048_incl_reduced_chi_both.pdf}\\
  \includegraphics  [width=9cm]{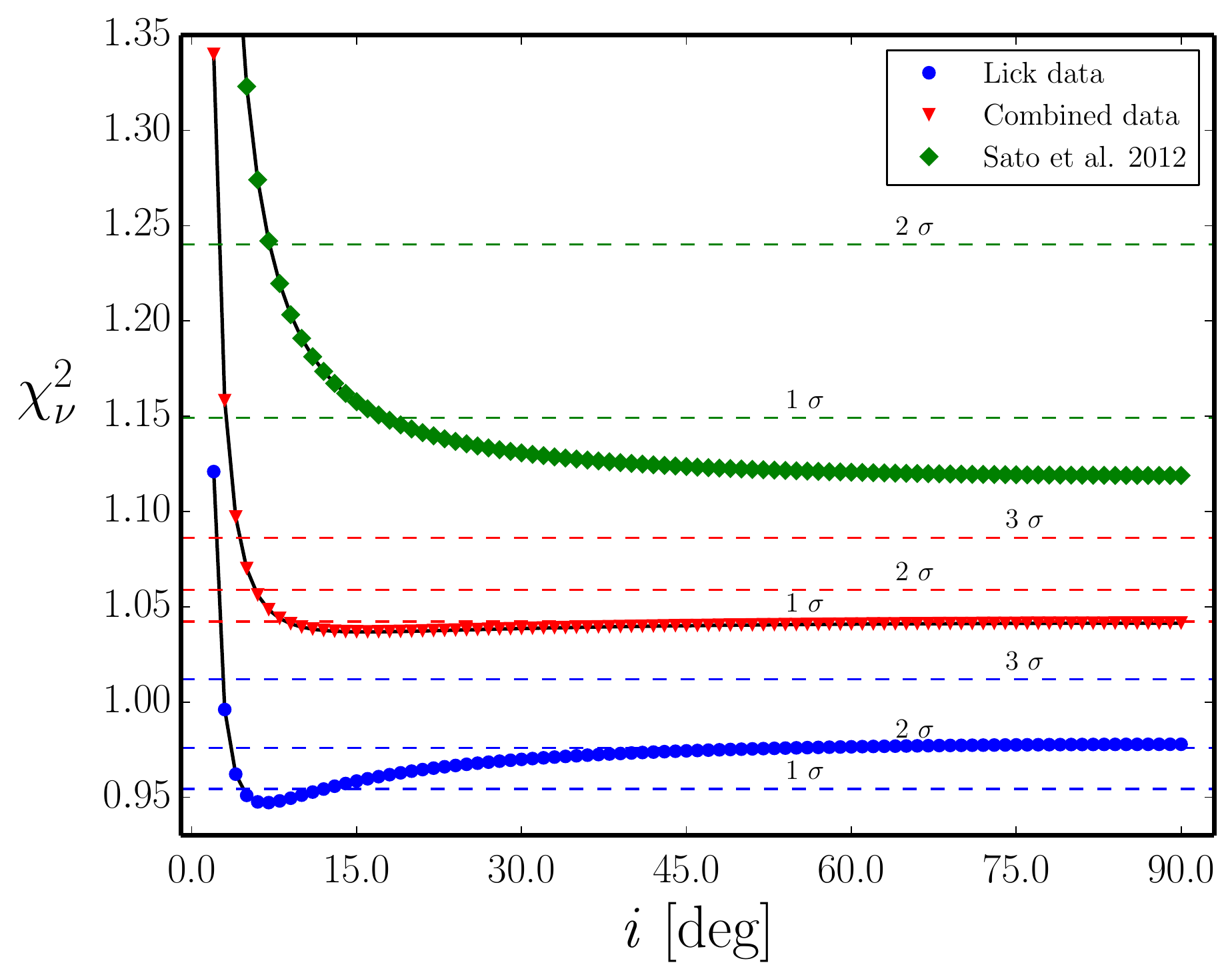}\\

  % \includegraphics  [width=9cm]{ HIP88048_incl_reduced_chi_combined_data.pdf}\\

 % \end{array}$
 \end{center}

  \caption{Resulting $\chi_{\nu}^2$ for coplanar dynamical fits as a function of inclination
 separately for Lick ({\it blue}) and OAO data ({\it green}), and for the combined data sets ({\it red}).
The test starts at $i = 90^\circ$ and goes down to $i = 3^\circ$ with a step size of $-1^\circ$.
Black lines are the 1\,D $\chi^2$ curves interpolated from the individual fits and applied to
individual data sets.
The dashed lines represent 1$\sigma$, 2$\sigma$ and 3$\sigma$ levels obtained from $\Delta\chi^2$
confidence values of 1.0, 4.0, and 9.0 larger than the $\chi^2$ minimum achieved for each data set.
%With $red$ is shown the same test, but using the combined RVs from Lick and OAO,
%while with $green$ is illustrated the same test applied only to the OAO data.
For the Lick data, the minimum at $i = 7^\circ$ is statistically significant at the 2$\sigma$ level, indicating that an inclined solution is
slightly preferred, while the OAO data has its minimum at $i = 90^\circ$,
showing that near edge-on configurations are preferred.
For the combined data the minimum is around $i = 16^\circ$, but is within 1$\sigma$ from
the coplanar edge-on fit.
   }
  \label{incl_fits}
  \end{figure}

\subsection{Coplanar inclined configurations}
\label{inclined}

For our coplanar inclined fitting we always fixed $\Delta \Omega = 0^\circ$ and set $i_1 = i_2 = i$.
In this way we keep the orbits in the same plane, and mutually inclined configurations are not allowed to occur.
While fitting we only alter the system's line of sight inclination $i$,
together with all other Keplerian parameters ($P$, $K$, $e$, $\omega$, $M_0$).
We obtained a best fit around $i = 16^\circ$, which has a slightly lower $\chi_{\nu}^2$ value when compared to the
edge-on case, but we also estimated very large inclination uncertainties for this fit.
This shows that there are only small, barely significant differences between coplanar edge-on and inclined dynamical fits, and we cannot constrain the orientation of the orbits from the current RV data. Thus, in our analysis we did not simply allow $i$ to vary as free parameter, but
we tested a set of coplanar fits as a function of inclination.
We started with our best edge-on dynamical fit at $i = 90^\circ$ (see Fig.~\ref{rv_plot} and Tab.~\ref{table:orb_par_stable})
and for the sequence of fits we adopted a step of $\Delta i = - 1^\circ$.
The minimum coplanar inclination we tested was at $i = 3^\circ$,
since a lower inclination leads to very massive companions approaching low-mass MS stars and above.
Moreover, we find that N-body models with very low inclinations %are totally inconsistent with the data
have large $\chi_{\nu}$ values (over 3$\sigma$ from the best fit), so there was no
reason to study  inclinations lower than $i < 3^\circ$.

Figure~\ref{incl_fits} illustrates the results from the coplanar inclined test.
The red dots represent the dynamical models obtained from the combined data set
and their $\chi_{\nu}^2$ values plotted versus the inclination $i$.
%Black line is the 1\,D $\chi_{\nu}^2$ surface interpolated from the fits position function of the inclination.
The red dashed lines represent 1$\sigma$, 2$\sigma$ and 3$\sigma$ levels obtained from
$\Delta\chi^2$ confidence values of 1.0, 4.0, and 9.0 larger than the $\chi^2$ minimum and scaled
accordingly by the number of the degrees of freedom ($DOF$) to match the $\chi_{\nu}^2$ levels in Fig.~\ref{incl_fits}.
Clearly, this test reveals that there is no significant $\chi_{\nu}^2$ minimum, although
slightly better fits can be obtained between $i \sim 10^\circ$ and 30$^\circ$.
All these fits, however, are within 1$\sigma$ from the coplanar
edge-on fit and as such the improvement can be considered to be insignificant.
Inclinations lower than $i \sim 10^\circ$ lead to models with rapidly increasing $\chi_{\nu}^2$ values,
and these are generally not consistent with the data.
We conclude that the dynamical fitting to the combined data is not
sensitive to inclinations larger than $i = 5^\circ$ (at 3$\sigma$), but such nearly face-on configurations are statistically strongly disfavored.
The global minimum appears to be at $i = 16^\circ$,
consistent with the best fit where we allowed the
system's line of sight inclination to be a free parameter.
The obtained orbital parameters and their errors from the
best coplanar inclined fit to the combined RV data
are given in Tab.~\ref{table:orb_par_stable}.

Despite the large companion masses obtained at $i = 16^\circ$,
this fit is stable for 1\,Gyr and it has a similar 6:1 resonant orbital evolution as
the coplanar edge-on case shown in Fig.~\ref{rv_plot}. The difference in this case, however, is that
all resonant angles evolve with higher frequency and larger libration amplitudes around 0$^\circ$.
The secular apsidal angle for this fit librates with a semi-amplitude of $34.9^\circ$, and
the resonant angles' libration semi-amplitudes are as follows:
$\Delta \theta_1 = 169.1^\circ$, followed by $\Delta \theta_2 = 134.4^\circ$,
$\Delta \theta_3 = 99.7^\circ$, $\Delta \theta_4 = 65.0^\circ$,
$\Delta \theta_5 = 32.2^\circ$, $\Delta \theta_6 = 49.2^\circ$.

Seeing that no confident constraints on the line of sight inclination can be made
based on the combined data, we were motivated to repeat the same test, but using Lick and OAO data, separately.
The idea was to see if the individual data sets contain any information
that can help us to define the inclination, and whether these sets are mutually consistent.
Figure~\ref{incl_fits} illustrates the same test applied to the Lick data in blue, and
in green results from the OAO data.
Individual dynamical models on both Lick and OAO data have different quality in terms of
$\chi^2$ as they have a different overall velocity precision (see Fig.~\ref{data_stat}).
Thus in Fig.~\ref{incl_fits} their $\chi_{\nu}^2$ curves and $\Delta\chi^2$ confidence levels
are above (OAO data) and below (Lick data) the statistics from the combined data set.
For the Lick data the minimum at $i = 7^\circ$ is significant at the 2$\sigma$ level indicating that an inclined solution is slightly
preferred, while the OAO data is not sensitive to inclination in the range $i = 20^\circ$ to $90^\circ$.

In summary, we conclude that the slight preference for a rather low inclination (near $i = 16^\circ$) is not statistically significant. The implied companion masses near or even above the hydrogen burning limit would in fact make the system even more perplexing, as discussed in Sect.~\ref{BDformation}.

%%%%%%%%%%%%%%%%%%%%%%%%%%%%%%%%%%%%%%%%%%%%%%%%%%%%%%%%%%%%%%%%%%%%%%%%%%%%%%%%%%%%%%
%
% \begin{table*}[!ht]
%
% %\resizebox{0.5\textheight}{!}
% %{\begin{minipage}{0.8\textwidth}
%
% \centering
% \caption{$\nu$\,Oph orbital constraints from Hipparcos}
% \label{table:astr_hipp}
%
%
% \begin{tabular}{ l cccccccccccc}
%
% \hline\hline  \noalign{\vskip 0.5mm}
%
%   % \multirow{2}{*}{Dataset} &
%     &&&&  \multicolumn{2}{c}{1 $\sigma$} &&
%       \multicolumn{2}{c}{2 $\sigma$} &&
%       \multicolumn{2}{c}{3 $\sigma$} \\
%   %          &&&&            1 &$\sigma$  & &   2 &$\sigma$    & &  3& $\sigma$       \\
%
%             &&  best-fit &&   min   & max    &&  min  &  max     &&   min &  max     \\
% \hline  \noalign{\vskip 0.9mm}
%
% $i_{\mathrm{b}}$ &&   117.52   &&   30.98 & 159.96 &&  20.33 & 164.29  &&   15.17 & 167.19    \\
% $i_{\mathrm{c}}$ &&   134.16   &&   10.21 & 171.97 &&  6.42  & 174.49  &&   4.81  & 176.03   \\
%
% \hline  \noalign{\vskip 0.9mm}
%
% $\Omega_{\mathrm{b}}$  &&    0.00   &&    101.64 &  265.04 && 163.95 & 164.92  &&  ? & ?    \\
% $\Omega_{\mathrm{c}}$  &&  127.65   &&      0.00 &  360.00 &&   0.00 & 360.00  &&  0.00? & 360.00?   \\
%
%  \hline\hline\
%
%
% \end{tabular}
% %\end{minipage}}
% \end{table*}
%

\section{Error estimation}
\label{Error}

\subsection{$\chi^2$ statistics}

\label{chi_error}

Error estimation plays an essential role in our orbital analysis, since it helps to
judge the reliability of the best-fit orbital parameters and to
reveal the orbital configurations that agree with the data within the statistical significance limits.
%Depending on the test we have performed to the data
%We use three different error estimators following the prescription in \citet{Press}.
%$\nu$\,Oph's orbital parameter errors from our individual best-fit coplanar models
%listed in Table \ref{table:orb_par_stable} are obtained directly form the $\chi^2$ fitting.
The individual best-fit parameter errors for $\nu$\,Oph in Tab.~\ref{table:orb_par_stable}
are obtained directly from the $\chi^2$ fitting
using the covariance matrix ($\sqrt{C_\mathrm{ii}}$). % implemented in the LM minimization algorithm.
These estimates represent symmetric 1$\sigma$ uncertainties of the best-fit parameters.
As can be seen from Tab.~\ref{table:orb_par_stable}, the $\nu$\,Oph system is very well
constrained with estimated orbital uncertainties usually below 1\%.
As discussed in Sect.~\ref{Orbital fit} we fit only the spectroscopic elements,
and thus the errors in physical parameters such as semimajor axes ($a_{1,2}$) and companion masses ($m_{1,2}$)
must be obtained through additional error propagation.
No errors in $i_{1,2}$ and $\Delta \Omega$ are obtained since
our dynamical model to the RV data was unable to provide an adequate constraint for these orbital parameters.
Instead, while fitting we always keep $i_{1,2}$ and $\Delta \Omega$ fixed.

%However, we demonstrated in Section \ref{inclined} and Fig.~\ref{incl_fits} that

Another method to estimate the orbital uncertainties from the RV data
is based on constant $\Delta\chi^2$ boundaries.
This method allows to explore the $\chi^2$ surface as a function of the fitting parameters, and as a result
an overall $\chi^2$ confidence statistics can be obtained \citep[see][]{Ford}.
The constant $\Delta \chi^2$ boundaries method is not practicable for models with
large number of free parameters (e.g.\ two-planet models),
due to the large computational resources needed to evaluate a smooth $\chi^2$ surface in the multi-dimensional
parameter space. However, if there are only one or two parameters of interest, the
$\Delta \chi^2$ technique can be a fast and valuable tool for estimating the parameter uncertainties.

For example, when there is only one free parameter, the values of $\Delta \chi^2$~=~1.0, 4.0, 9.0
larger than the global $\chi_{\nu}^2$ minimum will correspond to the 1$\sigma$, 2$\sigma$ and 3$\sigma$
confidence levels, while if there are two parameters, the $\Delta \chi^2$ values will be 2.3, 6.2 and 11.8.
As demonstrated in Sect.~\ref{inclined} and Fig.~\ref{incl_fits},
the $\Delta \chi^2$ boundaries technique can be used
effectively to constrain the significance of the line of sight
inclination $i$ for our dynamical models.
In Fig.~\ref{incl_fits} for each studied data set with $DOF$ degrees of freedom, we draw the 1$\sigma$, 2$\sigma$ and 3$\sigma$,
levels which correspond to
$\Delta \chi_{\nu}^2$ confidence values of $1.0 / DOF$, $4.0 / DOF$, and $9.0 / DOF$ larger than the $\chi_{\nu}^2$ minimum.
%The values of $\Delta\chi^2$~=~1.0, 4.0, 9.0 are the standard
%assuming only one fitting parameter is used (e.g. $i$).
We note that these $\Delta \chi^2$ boundaries provide the correct statistics only if
the applied model results in a minimum $\chi_{\nu}^2 = \chi^2/ DOF = 1$.
None of our $\chi_{\nu}^2$ extrema for the three cases shown in Fig.~\ref{incl_fits}
are actually exactly at unity, but they are close enough for considering the confidence levels as valid.

\subsection{Bootstrap sampling}
\label{boot_error}

Apart from the covariance matrix and the $\Delta \chi^2$ statistics,
it is possible to carry out an independent error estimation, and to validate
the uncertainties of the orbital elements, by applying a bootstrap resampling of the original RV
data. This method uses synthetic data sets, each of which can be fitted with a
Keplerian or dynamical model, so that an overall statistical distribution of the
fitted parameters can be obtained.

\begin{figure*}[tp]
 \begin{center}$
% \begin{array}{ccc}
%    \includegraphics  [width=15.4cm]{\pathh  hip88048/bootstrap_10k_all_opt_1/Planet_1_VE_10k_edgeon_shared.pdf} \\
%    \includegraphics  [width=15.4cm]{\pathh  hip88048/bootstrap_10k_all_opt_1/Planet_2_VE_10k_edgeon_shared.pdf} \\
%    \includegraphics  [width=15.4cm]{\pathh  hip88048/bootstrap_10k_all_opt_1/theta_boot.pdf} \\
%   \end{array}$
%  \end{center}
 \begin{array}{ccc}
    \includegraphics  [width=17.5cm]{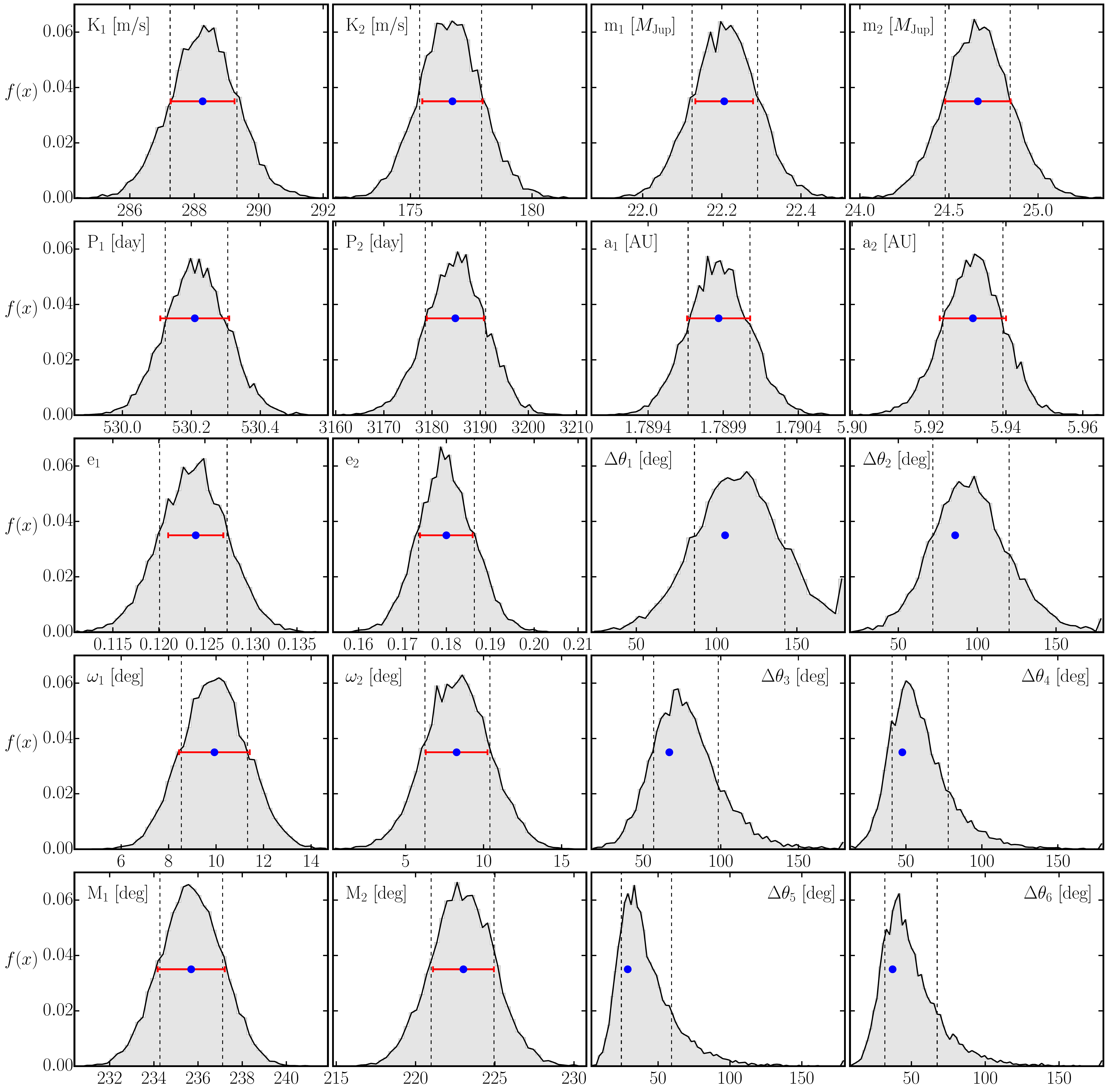} \\
  \end{array}$
 \end{center}

 %K_phase_dist.pdf

  \caption{The first two columns show the sampling distribution of all
  fitting parameters ($K$, $P$, $e$, $\omega$, $M$) for a coplanar and edge-on
  configuration constructed from 5000 bootstrap samples. All bootstrap samples are stable for 10\,Myr.
  The third and fourth columns show the derived distribution of companion masses ($m$) and semimajor axes ($a$), and
  the libration semi-amplitudes of the resonance angles ($\theta_{1}, \ldots, \theta_{6}$).
  The blue dots are the location of the best dynamical fit in phase space
  for both companions. The red error bars are the uncertainties estimated from the covariance matrix, while
  the errors in $a$ and $m$ are obtained trough error propagation from these uncertainties.
  Clearly, for all orbital elements the best-fit values and their errors from the covariance matrix are consistent
  with the bootstrap distribution peak and the corresponding 68.3\% confidence level (vertical dashed lines).
  %All bootstrap samples are stable for 10 Myr.
 % Third and fourth columns show the derived distribution of companion mass and semimajor axes and
 % the resonance angles ($\theta_{1,6}$) libration amplitudes distribution.
}
  \label{bootstrap_copl}
  \end{figure*}

We followed the bootstrap prescription described in \citet{Press} and
\citet{Tan2013}. Briefly, for each data set containing $N$ data points, we generated 5000
synthetic samples containing $N$ data points, chosen randomly from the
original
data set with replacement. The OAO and Lick data sets were re-sampled
separately and then combined. To each alternative combined data
set we fitted a two planet dynamical model and tested for stability as
defined in Sect.~\ref{Orbital fit}.
The sampling distribution of the fitted orbital parameters
obtained with the bootstrap method for the edge-on coplanar configuration
are illustrated in Fig.~\ref{bootstrap_copl}.
We also show the distribution of the derived semimajor axes
($a_{1}, a_{2}$) and the companion masses ($m_{1}, m_{2}$).
Since all simulated fits appeared to be stable ($t_{\mathrm{max}} = 10$\,Myr),
Fig.~\ref{bootstrap_copl} also shows the distribution of the
libration semi-amplitudes for all six resonance angles ($\theta_1, \ldots, \theta_6$).

In Fig.~\ref{bootstrap_copl} we also provide a comparison between the errors estimated from the
covariance matrix of the best fit to the original data, and the
$1\sigma$ confidence interval from the bootstrap distribution.
In all plots, blue dots represent the best-fit values obtained from the best
dynamical fit, while red error bars are the estimated uncertainties from
the covariance matrix (see Tab.~\ref{table:orb_par_stable}).
We find that all best-fit parameters and their errors are consistent
with the bootstrap distribution peak and the 68.3\% confidence level
(vertical dashed lines on Fig.~\ref{bootstrap_copl}). The bootstrap
distribution is very symmetrical for all fitted parameters, and can be approximated well with a normal distribution.

\section{Dynamical analysis}
\label{Stability}

The statistical and dynamical properties of the fits within the statistically permitted region of the parameter space around the best fit can be obtained
using a systematic $\chi_{\nu}^2$ grid-search technique coupled with dynamical fitting, as was
previously demonstrated in \citet{Lee2006}, \citet{Tan2013}, and \citet{Trifonov2014}.
To study the possible orbital configurations for the $\nu$\,Oph system, we select pairs of parameters and construct high-density 2D grids consisting of $50 \times 50$ points for each of them. For each point on the grid, we keep these two parameters fixed, and perform a $\chi_{\nu}^2$ minimization with a dynamical model, allowing all other parameters to vary. By performing $50 \times 50$ dynamical fits, we thus obtain $\chi_{\nu}^2$ contours in the plane of the two chosen grid parameters. From these, we derive 1$\sigma$, 2$\sigma$ and 3$\sigma$ confidence levels based on the $\Delta \chi^2$ statistics.
As a final step, we integrate these fits for 10\,Myr with {\it SyMBA} to study their stability and dynamical evolution.

\begin{figure*}[tp]
 \begin{center}$
\begin{array}{ccc}

  \includegraphics  [width=18cm]{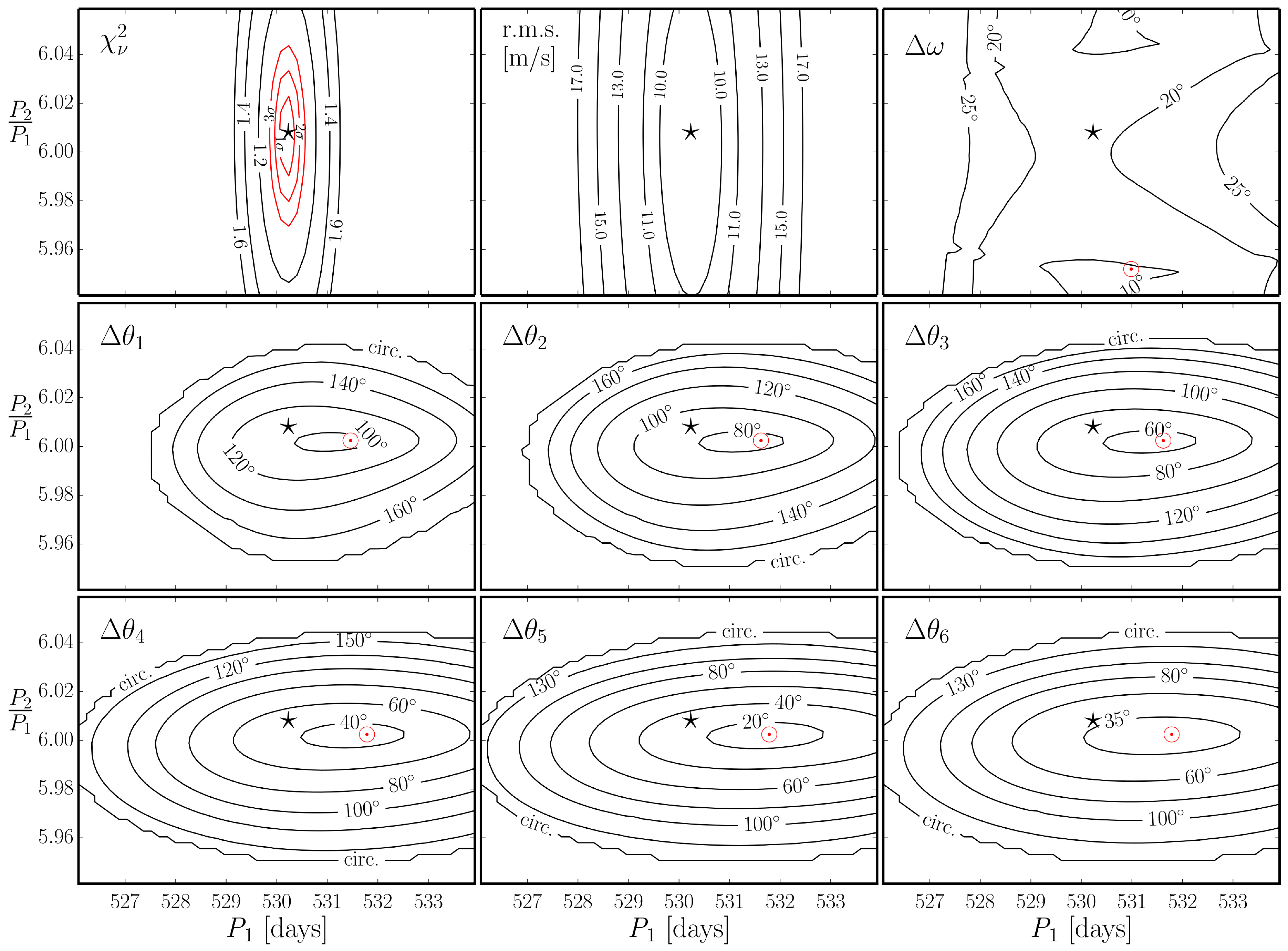} \\

  \end{array}$
 \end{center}

  \caption{Results from the $P_2 / P_1$ versus $P_1$ grid constructed from coplanar edge-on dynamical fits.
  The separate panels are self-explanatory: The top left panel shows the $\chi_{\nu}^2$ grid surfaces, where the red contours denote the
  1$\sigma$, 2$\sigma$ and 3$\sigma$ levels from the grid's best fit. The best fit itself is
  found near the 6:1 period ratio and is marked with a black star symbol in all panels.
  The radial velocity r.m.s.\ contour levels show consistency with the $\chi_{\nu}^2$ surface, with lower
  r.m.s.\ values found around the best fit. The other panels show all the 6:1 MMR
  resonance angles (the secular $\Delta \omega$ and $\theta_{1}, \ldots, \theta_{6}$), and their libration semi-amplitudes on the grid.
  The red $\odot$ symbol marks the minimum libration amplitude for each resonant angle.
  These plots show that almost all fits within the formal 3$\sigma$ confidence level exhibit
  resonance behavior with all six 6:1 MMR angles librating. See text for details.
}
  \label{p2p1_i90}
  \end{figure*}

   \begin{figure*}[htp]
 \begin{center}$
\begin{array}{ccc}

    \includegraphics  [width=18cm]{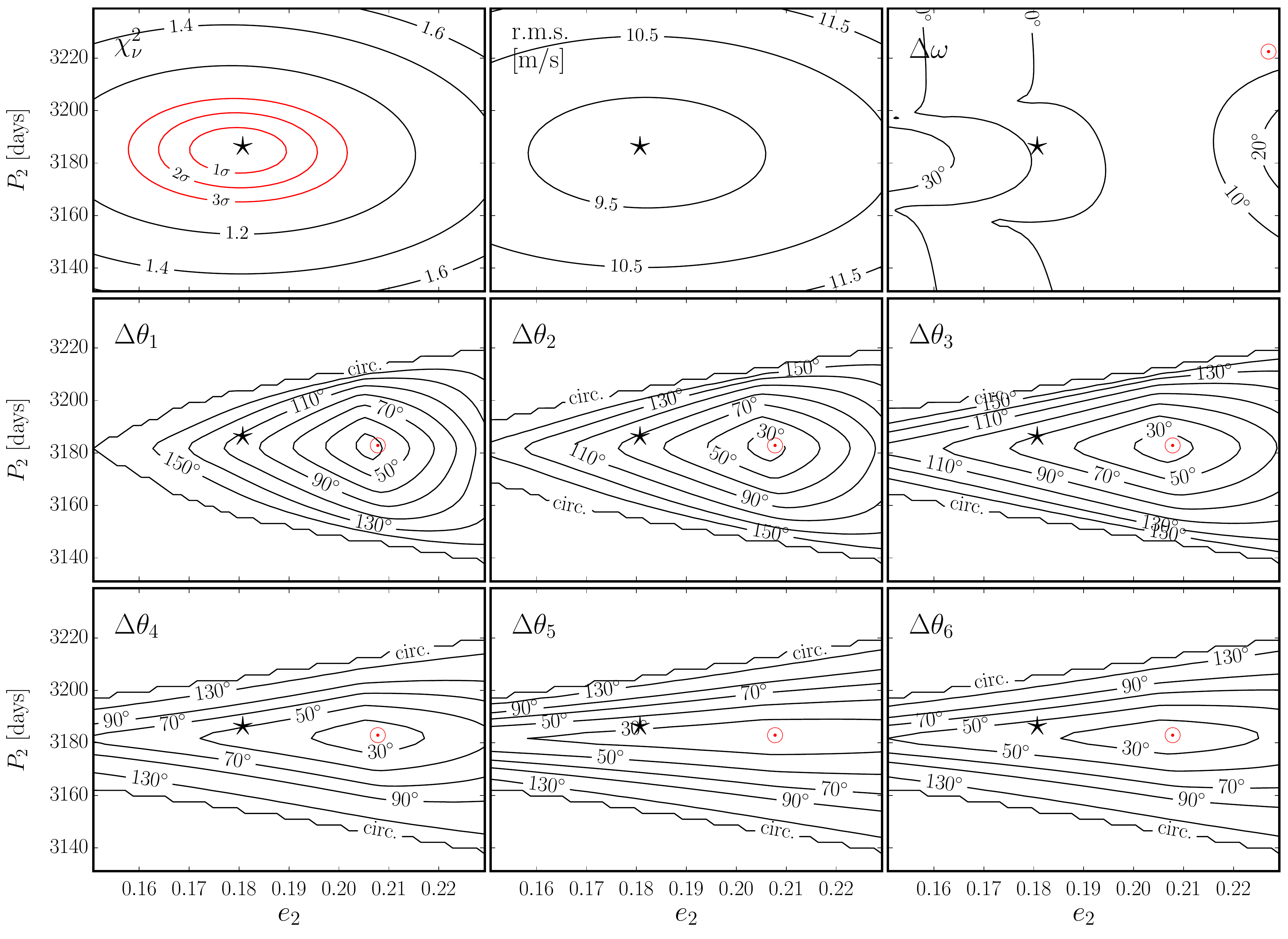} \\

  \end{array}$
 \end{center}

   \caption{Results from the $P_2$ versus $e_2$ grid constructed from coplanar edge-on dynamical fits. The nine panels show contours for the same quantities as those in Fig.~\ref{p2p1_i90}.
   }

  \label{p2e2_i90}
  \end{figure*}

   \begin{figure*}[htp]
 \begin{center}$
\begin{array}{ccc}

  \includegraphics  [width=18cm]{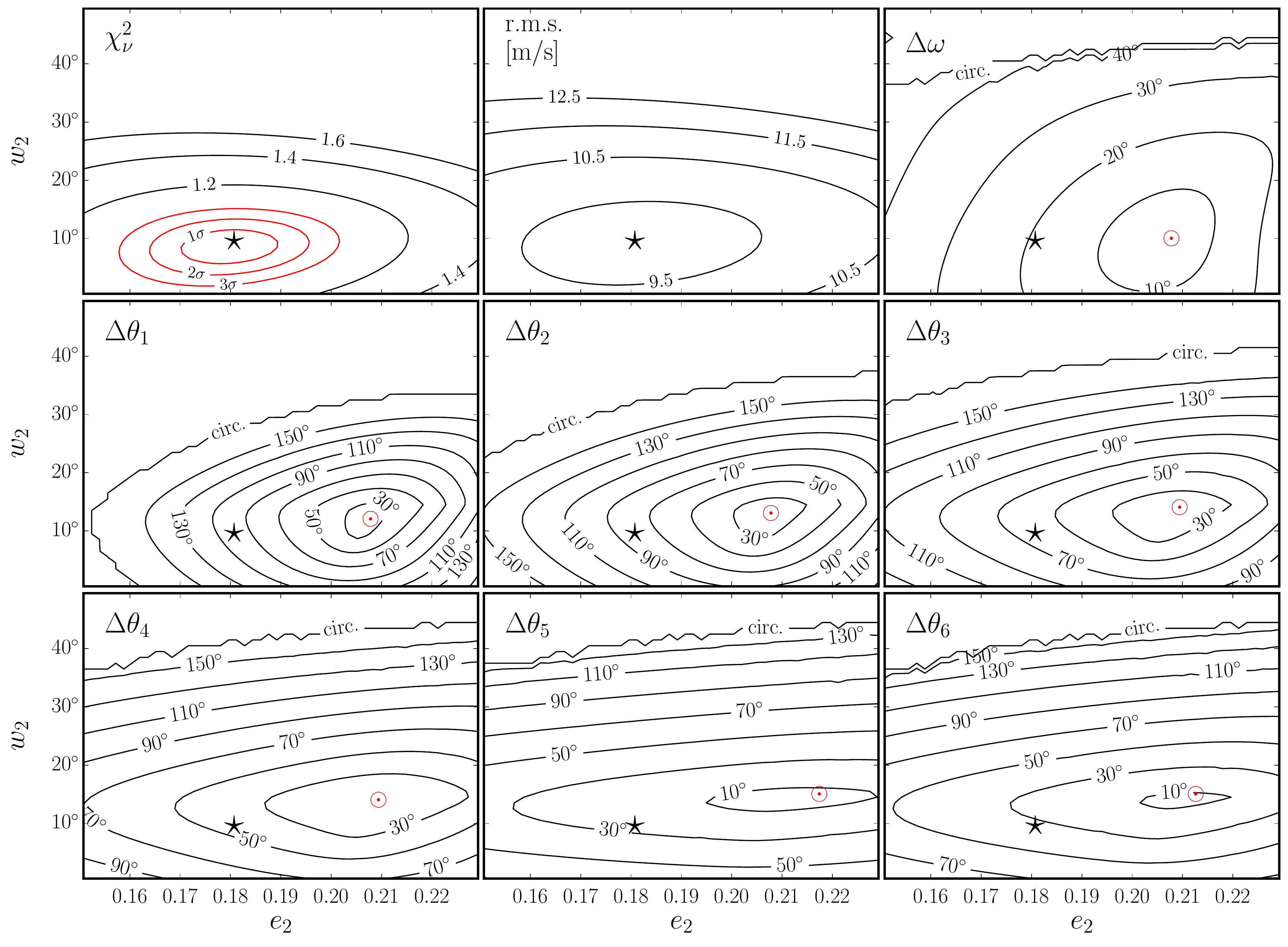} \\

  \end{array}$
 \end{center}

   \caption{As in Fig.~\ref{p2p1_i90} and Fig.~\ref{p2e2_i90}, but for $e_2$ versus $\omega_2$.
 All fits on this grid are stable for at least 10\,Myr, and the fits within the formal 3$\sigma$ confidence level are in 6:1 MMR
  with $\theta_{1}, \ldots, \theta_{6}$ librating around 0$^\circ$. See text for details.
}

  \label{e2w2_i90}
  \end{figure*}

\subsection{Coplanar configurations}
\label{dyn_prop}

In an attempt to better understand the resonant nature of the $\nu$\,Oph system,
and to see if the resonance configuration of the system is preserved
across the orbital phase space allowed by $\chi_\nu^2$, we constructed three different 2D coplanar edge-on grid combinations.
We start with a grid, where we fix the period ratio $P_2 / P_1$ and the period of the inner planet $P_1$,
and then we construct $P_2$ vs.\ $e_2$ and $\omega_2$ vs.\ $e_2$ grids.
For the first, we systematically vary $P_1$ in the range between 526\,d and 534\,d,
while varying $P_2$ in such way that for each $P_1$ the resulting $P_2 / P_1$
increases from 5.95 to 6.05 with a constant step of 0.002.
We test the $P_2$, $e_2$ grid for $P_2 = 3130$\,d to 3240\,d and
the $\omega_2$, $e_2$ grid for $\omega_2 = 0^\circ$ to 50$^\circ$; for both
grids $e_2$ was varied in the range $e_2 = 0.15$ to 0.23.
For all grids we examine the statistical properties around the best achieved fit,
the stability and orbital evolution, and we record the distribution of all orbital elements and
the libration amplitudes.

Results from the $P_2 / P_1, P_1$ grid are shown in Fig.~\ref{p2p1_i90},
while results from the $P_2, e_2$ and $e_2, \omega_2$ grids are shown in
Fig.~\ref{p2e2_i90} and Fig.~\ref{e2w2_i90}, respectively.
These figures show the achieved $\chi_{\nu}^2$ and the radial velocity r.m.s.\ surface from these tests,
along with the dynamical properties around the best fit in terms of
the derived libration semi-amplitudes for
$\Delta \omega$ and all six resonant angles $\theta_{1}, \ldots, \theta_{6}$.
We marked the best fit in the grid with a star, while red contours trace the
1$\sigma$, 2$\sigma$ and 3$\sigma$ significance levels in the grid
obtained from $\Delta \chi^2$ statistics. We note that in all cases the chosen parameter ranges are sufficiently large for the grids to encompass the full 3$\sigma$ contours.

We find that all studied combinations lead to similar conclusions, and thus we summarize them as follow:
All studied fits are stable for 10\,Myr, and therefore no stability borders exist on these grids.
The best fits found on the individual grids (black star symbol)
are near the 6:1 period ratio and exhibit very similar orbital parameters and evolution
when compared with the best coplanar edge-on dynamical fit shown in Tab.~\ref{table:orb_par_stable}.
Small differences occur in the fixed parameters, but that is expected given the finite resolution
of the examined grids. We conclude that the best fits for all three grids are associated with the
best coplanar edge-on fit for $\nu$\,Oph and no other local $\chi_{\nu}^2$ minima exists
on the studied grids. The radial velocity r.m.s.\ contours are consistent with the $\chi_{\nu}^2$ surface.
We see lower r.m.s.\ values around the best fit, and the r.m.s.\ level
smoothly increases between the best fit and the 3$\sigma$ level.
From these grids we find that all the 6:1 MMR resonance angles librate within the
3$\sigma$ level and beyond, meaning that all significant fits in these grids are in 6:1 MMR.
Figure~\ref{p2p1_i90}, Fig.~\ref{p2e2_i90} and Fig.~\ref{e2w2_i90} show
that the resonance angle semi-amplitudes at the best fit
are in very good agreement with those derived from the orbital evolution of
the best coplanar edge-on fit illustrated in Fig.~\ref{theta_evol}.
It is clear, however, that all resonance angles have a minimum in libration amplitude (red $\odot$ symbol
on the figure subplots) different form the best fit.
This amplitude minimum appears to be on the exact 6:1 period ratio, but over 3$\sigma$ away from our best
fit. This means that a deeper resonance state does exist for this system, but is not consistent with the data.

For many fits in the outer regions of the grids all resonance angles circulate, and thus these fits
are not associated with a MMR behavior.
Nevertheless, looking at the subplots for $\Delta \omega$ in all three grid combinations,
one can see that $\Delta \omega$ actually almost never circulates, and the system
remains in an aligned geometry.
This result suggests that these phase space regions are not random,
but the system is involved in secular interactions where $\Delta \omega$ librates around 0$^\circ$.
We find that the libration amplitude of $\Delta \omega$ depends
on the initial $\Delta \omega$ from which we start the numerical integrations.
In the case of $P_2 / P_1, P_1$ and $P_2, e_2$ grids $\omega_1$ and $\omega_2$
are floating, and usually the best fit suggests nearly aligned orbital geometry.
%The later orbital evolution of these fits shows that the system's $\Delta \omega$ keeps
%librating around 0$^\circ$, while all resonant angles circulate.
In the case of the $e_2, \omega_2$ grid, $\omega_2$ is fixed between 0$^\circ$ and 50$^\circ$,
and the initial $\Delta \omega$ can reach $\gtrsim 30^\circ$.
For such initial values we find that $\Delta \omega$ circulates.

However, the grid regions where all the resonance angles circulate
have very large $\chi_{\nu}^2$ values, and thus such fits are statistically very unlikely
to explain the $\nu$\,Oph RV data.
It is nevertheless interesting to investigate whether these fits are long-term stable over $\sim 1$\,Gyr.
We test a number of these cases for each grid in the region where the $\chi_{\nu}^2$ is larger
(usually at the grid corners), and we find that even non-MMR orbital configurations are stable for 1\,Gyr.
We conclude that a large region of parameter space around the best fit contains only stable configurations, so that no stability constraints on the possible orbital configurations can be obtained for edge-on and coplanar orbits.

 %(TBD more more more)

We repeated the $P_2 / P_1$ vs.\ $P_1$, $P_2$ vs.\ $e_2$, and $e_2$ vs.\ $\omega_2$ grids
for inclined coplanar configuration where we set $i = 30^\circ$.
In this way we obtained approximately a factor of two times more massive companions for each fit.
We find, however, that doubling the companion masses has little influence on
$\chi_{\nu}^2$, and the radial velocity r.m.s.\ grid contours resemble those for $i = 90^\circ$.
This result is not very surprising given the results presented in Fig.~\ref{incl_fits}.
%We find earlier that for the combined RV data there is no really a significant qualitative difference
%in the fits between $i$~=~30$^\circ$ and $i$~=~90$^\circ$.
Similar to the edge-on case the $\chi_{\nu}^2$ minimum on all grids matches with the
best fit for $i = 30^\circ$ from the coplanar inclined test performed in Sect.~\ref{inclined}.

The main goal for this test was to examine the resonance state of the $\nu$\,Oph system
assuming more massive bodies in orbit.
We find that the test carried out for $i = 30^\circ$ is almost identical with
the edge-on case. All grid points led to stable solutions, and the fits within the $3\sigma$ confidence
contours are all in 6:1 MMR. The resonant regions on the grids had somewhat smaller
surface area when compared to the edge-on case, but they exhibit similar libration amplitudes,
while the libration frequency is higher as can be expected for more massive interacting bodies.

These results show that the $\nu$\,Oph system is deeply trapped in a 6:1 MMR, and that this configuration
is dynamically possible for a relatively large range of companion masses.

%
%  \begin{figure*}[htp!]
%  \begin{center}$
% \begin{array}{ccc}
%
%   \includegraphics  [width=6cm]{i1i2_mvs_0.pdf}
%   \includegraphics  [width=6cm]{i1i2_mvs_30.pdf}
%   \includegraphics  [width=6cm]{i1i2_mvs_60.pdf} \\
%   \includegraphics  [width=6cm]{i1i2_mvs_90.pdf}
%   \includegraphics  [width=6cm]{i1i2_mvs_120.pdf}
%   \includegraphics  [width=6cm]{i1i2_mvs_150.pdf} \\
%   \includegraphics  [width=6cm]{i1i2_mvs_180.pdf} \\
%
%
%   \end{array}$
%  \end{center}
%
%   \caption{ Mutually inclined grids for different $\Delta\Omega$. The red contours illustrate the  1$\sigma$, 2$\sigma$ and 3$\sigma$ confidence levels,
%   while the black dashed contours are the mutual inclination bothers. The grids stability is tested for 1 Myr and gray filled contours show the grid areas where
%   the orbits are stable. See text for details.}
%   \label{mut_incl}
%   \end{figure*}
%

%\begin{figure*}[htp!]
% \begin{center}$
%\begin{array}{ccc}
%
%  \includegraphics  [width=6cm]{i1i2_mvs_0a.pdf}
%  \includegraphics  [width=6cm]{i1i2_mvs_30a.pdf}
%  \includegraphics  [width=6cm]{i1i2_mvs_60a.pdf} \\
%  \includegraphics  [width=6cm]{i1i2_mvs_90a.pdf}
%  \includegraphics  [width=6cm]{i1i2_mvs_120a.pdf}
%  \includegraphics  [width=6cm]{i1i2_mvs_150a.pdf} \\
%  \includegraphics  [width=6cm]{i1i2_mvs_180a.pdf}
%  \includegraphics  [width=6cm]{i1i2_mvs_210.pdf}
%  \includegraphics  [width=6cm]{i1i2_mvs_240.pdf} \\
%  \includegraphics  [width=6cm]{i1i2_mvs_270.pdf}
%  \includegraphics  [width=6cm]{i1i2_mvs_300.pdf}
%  \includegraphics  [width=6cm]{i1i2_mvs_330.pdf} \\
%
%
%  \end{array}$
% \end{center}
%

 \begin{figure*}[htp!]
 \begin{center}$
  \includegraphics  [width=18cm]{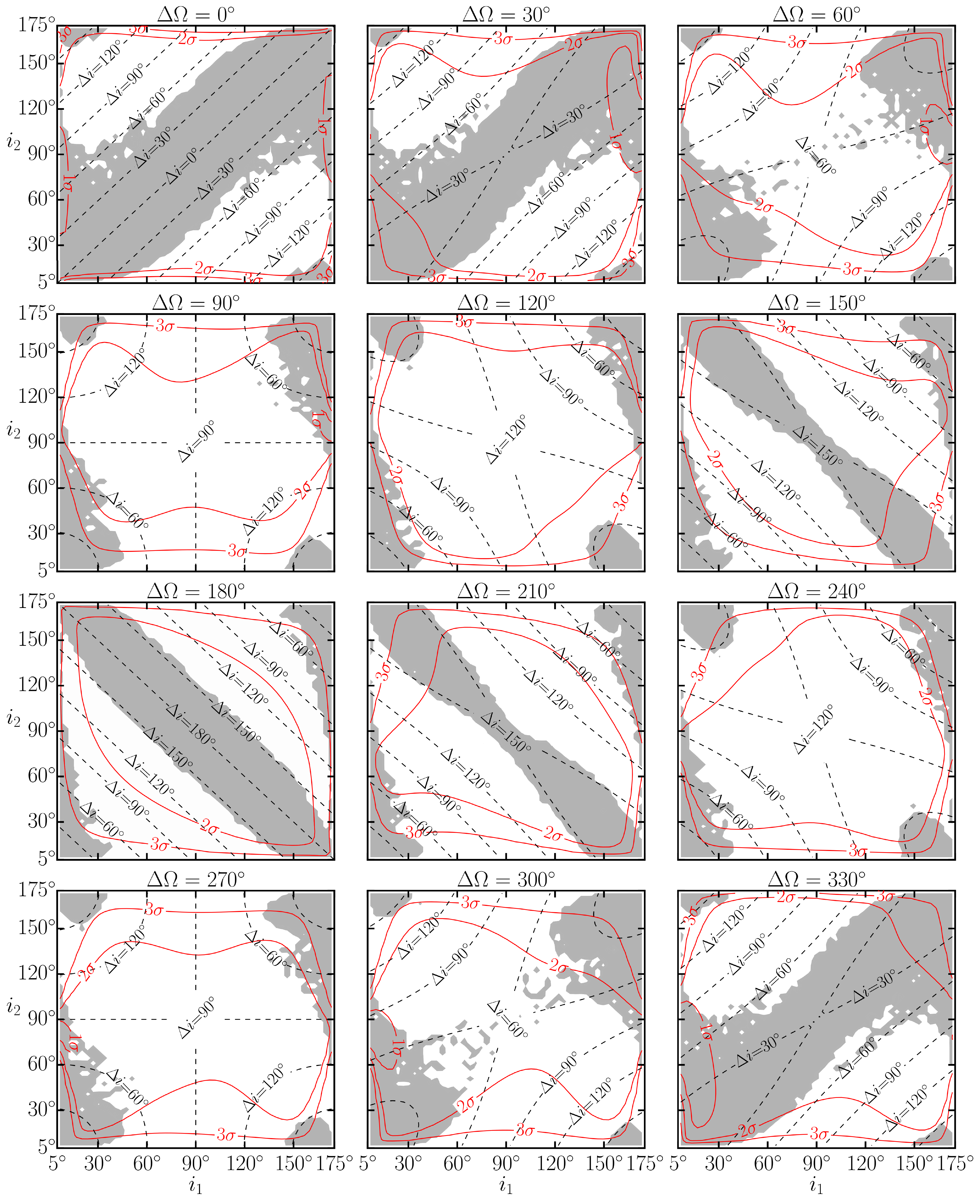}$
 \end{center}

  \caption{Mutually inclined grids for different $\Delta \Omega$. The red contours illustrate the 1$\sigma$, 2$\sigma$ and 3$\sigma$ confidence levels,
  while the black dashed lines delineate constant values of the initial mutual inclination $\Delta i$. The stability of the best-fit solution for each grid point is tested for 1\,Myr, and gray filled contours show the grid areas where the orbits are stable. See text for details.}
  \label{mut_incl}
  \end{figure*}

\subsection{Mutually inclined configurations}
\label{mut_inclined}

We investigated the Hipparcos Intermediate Astrometric Data for $\nu$\,Oph in an attempt to set
constraints on the inclinations and the ascending nodes of the system as was demonstrated in \cite{Reffert2011}.
We found that all but the lowest inclinations (down to about $i_{1,2} = 5^{\circ}$) for both companions
are consistent with the Hipparcos data, while $\Omega_1$ and $\Omega_2$ are practically unconstrained.
Therefore, no meaningful constraints on the orbital configuration could be derived from the Hipparcos astrometry.

Although our dynamical fits to the RV data are also unable to constrain the orbital orientation,
we test a large number of mutually inclined configurations by constructing $i_1$ vs.\ $i_2$ grids.
These grids are made in the range between $i_{1,2} = 5 \ldots 175^\circ$,
meaning that each step corresponds to $\delta i_{1,2} = 3.4^\circ$.
The mutual inclination, however, depends also on the difference between
the orbital ascending nodes $\Delta \Omega = \Omega_1 - \Omega_2$:
\begin{equation}\label{delta_i}
  \cos \Delta i = \cos i_1 \cos i_2 + \sin i_1 \sin i_2 \cos \Delta \Omega~~.
\end{equation}
Therefore we create a total of twelve $i_1, i_2$ grids with fixed values for
$\Delta \Omega = 0^\circ$, 30$^\circ$, 60$^\circ$, \ldots, 330$^\circ$.
These grids cover a large set of mutually inclined configurations
in the range $\Delta i = 0 \ldots 180^\circ$; from coplanar edge-on to
highly inclined and retrograde orbits.
For each set of $i_1, i_2$ and $\Delta \Omega$ we apply our dynamical fitting routine
and collect the $\chi_{\nu}^2$ value and the best fit orbital elements.
For all fits obtained on these grids we test the long-term orbital evolution for 1\,My.

Figure~\ref{mut_incl} visualizes the results from our $i_1, i_2$ vs.\ $\Delta \Omega$ grids.
The red contours show the 1$\sigma$, 2$\sigma$ and 3$\sigma$ levels
measured form the global best-fit model on the  $i_1 , i_2$ vs.\ $\Delta \Omega$ grids.
The dashed contours mark the initial mutual inclination, in steps of $\Delta i = 30^\circ$,
obtained with Eqn.~\ref{delta_i} from $i_1$, $i_2$ and the given $\Delta \Omega$.
Gray filled regions show the fits stable for at least 1\,Myr.

For the grid with $\Delta \Omega$ fixed at 0$^\circ$, the mutual inclination
only depends on $\Delta i = | i_1 - i_2|$, and therefore all coplanar inclined configurations are
located on the diagonal with $i_{1} = i_{2}$. %~=~5$^\circ$ and $i_{1,2}$~=~175$^\circ$.
The fits located on this diagonal are a repetition of the test performed
in Sect.~\ref{inclined} and shown in Fig.~\ref{incl_fits} for the combined RV data set.
In agreement with the results presented in Sect.~\ref{inclined}, the coplanar inclined fits on this grid are all stable.
The same is true for a large fraction of mutually inclined fits with $\Delta i \lesssim 45^\circ$.
The best stable fits located within the 1$\sigma$ confidence region on the grid, however, are
located at $i_1 \lesssim 15^\circ$ or $i_1 \gtrsim 165^\circ$ (i.e., $\sin i_1 \lesssim 0.26$), where the inner companion is in the low-mass stellar range.
This stable region covers a more extended range of mutual inclinations
from configurations with $\Delta i \approx 30^\circ$ and large companion masses to about
$\Delta i \approx 100^\circ$, where the outer companion has a mass close to its minimum, while
the inner companion mass is near its maximum on the grid.
Additionally, two small stable regions exist for inclined configurations
with $\Delta i \gtrsim 150^\circ$ (which is equivalent to $\Delta i \lesssim 30^\circ$ with retrograde orbits),
but these are over 2$\sigma$ away from the best fit.
The large S-shape stable island for $\Delta \Omega = 0^\circ$
gives a large range of companion masses and mutually inclined
configurations that can explain the $\nu$\,Oph system rather well.

The case for $\Delta \Omega$ fixed at 30$^\circ$ shows a similar stable
region as the case of $\Delta \Omega = 0^\circ$, but with $\chi_{\nu}$ confidence contours
suggesting decreasing quality of the fits in the stable region.
Only one well-defined 1$\sigma$ region is found in this grid, and almost all fits within it are stable.
These fits are located at $i_1 \gtrsim 150^\circ$ covering $\Delta i$ between 10$^\circ$ and 90$^\circ$.
All other fits in the stable S-shape region are more than 1$\sigma$ away from the best fit.

For $\Delta \Omega = 60^\circ$, 90$^\circ$ and 120$^\circ$, the grids contain only configurations with high mutual inclinations, some with mostly prograde, others with mostly retrograde orbits. This has a major impact on the stable regions and the confidence contours.
For these three grids the stable solutions are mostly
located near the corners, corresponding to high-mass companions with relatively
small prograde or retrograde mutual inclinations.
The quality of these fits indicates that almost all of them are not within the 1$\sigma$ region.
For $\Delta \Omega = 60^\circ$, the 1$\sigma$ contours are rather similar
to those at $\Delta \Omega = 30^\circ$, but the central part of this grid has only few
stable fits, which in fact will most likely turn out to be unstable if tested for more than 1\,Myr.
The grids for $\Delta \Omega = 90^\circ$ and $\Delta \Omega = 120^\circ$ consist mostly of nearly perpendicular geometries, and the fits
in the central 2$\sigma$ regions are nearly all unstable. The large unstable regions at high mutual inclination in these and the other panels are most likely due to Kozai-Lidov cycles, which seem to affect the system stability.

A large stable region begins to emerge at very high mutual inclinations ($\Delta i \gtrsim 150^\circ$) at $\Delta \Omega = 150^\circ$.
For $\Delta \Omega = 180^\circ$, this stable region is well defined around the coplanar and retrograde diagonal at $\Delta i = 180^\circ$.
The fits at the central part for $\Delta \Omega =150^\circ$ and 180$^\circ$ are close to the 1$\sigma$ confidence level.
Clearly for retrograde orbits the system is only stable within relatively small limits of $\lesssim 30^\circ$ for the mutual inclination (i.e., $\Delta i \gtrsim 150^\circ$).

The panels in Fig.~\ref{mut_incl} for $\Delta \Omega = 210^\circ \ldots 330^\circ$ are very similar to those for $\Delta \Omega = 30^\circ \ldots 150^\circ$ (in reverse order). This is expected, as the transformation ($i_{1,2} \rightarrow 180^\circ - i_{1,2}$, $\Delta \Omega \rightarrow - \Delta \Omega$) leads to a system that is observationally indistinguishable and has the same stability properties. The small differences between the corresponding panels are due to numerical noise.

If the $\nu$\,Oph system has indeed mutually inclined orbits,
then most likely the system exists with relatively
low mutual inclinations in prograde orbits,
or even lower mutual inclination in retrograde orbital motion.
Except for very small areas of parameter space, mutual inclinations with $\Delta i$ between 60$^\circ$ and 150$^\circ$
lead to instability on very short time scales, due to the Kozai-Lidov effect.
We caution, however, that the apparently stable regions might become smaller if longer integrations are carried out for the stability tests.

\section{Discussion}
\label{Discussion}

\subsection{The nature of the Brown Dwarf Desert}

In 2003 the Working Group on Extrasolar Planets of the International Astronomical Union adopted a working definition according to which a substellar companion to a star is to be called ``planet'' if its mass is less than $13\,M_{\mathrm{Jup}}$, and ``brown dwarf'' if it is higher \citep{IAU2007}. The main purpose of this distinction is the introduction of an unambiguous nomenclature that is closely linked to an observable quantity, although in the case of radial-velocity measurements the $\sin i$ ambiguity remains. The boundary at $13\,M_{\mathrm{Jup}}$ is motivated by the deuterium burning limit, but from the point of view of companion formation mechanisms it is completely arbitrary.

Nevertheless, the distinction between planets and brown dwarf companions took on a second meaning with the realization that very few $\sim 1\,M_\odot$ main-sequence stars harbor companions in the range 5 to $80\,M_{\mathrm{Jup}}$, with orbital periods up to a few years \citep{Marcy2000}. This ``Brown Dwarf Desert'' can be understood as a deep minimum in the companion mass function between the planetary and stellar mass ranges \citep{Grether2006}, related to different formation mechanisms: Whereas binary stars form in a cloud fragmentation process favoring pairs with nearly equal masses and thus a companion mass function with a positive slope, planets form in circumstellar disks, with a mass function with negative slope over the range considered here (gas giants with $m \gtrsim 1\,M_{\mathrm{Jup}}$). In this picture, the location of the minimum between the two companion mass functions in the range between the deuterium and hydrogen mass burning limits is coincidental, but it provides a tentative connection between the mass-based nomenclature and the putative formation channels.

The large radial-velocity exoplanet surveys carried out during the past decades have also discovered a fair number of companions with $m \sin i$ in the brown dwarf range \citep[e.g.,][]{Nidever2002, Patel2007, Sahlmann2011, Wilson2016}. For some of these objects it has been possible to detect the astrometric signature of the orbital motion in the intermediate data of the {\it Hipparcos} mission, and thus to measure $\sin i$. In most cases, this has led to the realization that the secondaries are low-mass stars in nearly face-on orbits \citep[e.g.,][]{Halbwachs2000, Sahlmann2011, Wilson2016}, but a few true brown dwarf companions have also been confirmed in this way \citep{Sozzetti2010, Reffert2011}. Radial-velocity follow-up of transiting Jupiter-size objects is an alternative way to firmly establish brown dwarf companions, as in the case of CoRoT-3\,b \citep{Deleuil2008}.

Brown dwarf candidates have also been discovered in orbits around late G and K giants \citep[e.g.,][]{Liu2008, Mitchell2013, Reffert2015}, in addition to the two objects that are the subject of this paper. As the giant star surveys probe stellar masses up to $\sim 4\,M_\odot$, they may help to answer questions about the properties of the Brown Dwarf Desert. One might for instance suspect that the characteristic mass of the Desert (i.e., the mass at which the planetary and stellar companion mass functions intersect) increases with the mass of the host star. This would be expected if the masses of planet-forming circumstellar disks increase with the stellar mass \citep[e.g.,][]{Pascucci2016}, and if the occurrence rate of stellar binaries depends on the mass ratio $q$ rather than the absolute mass of the secondary (\citealt{Chabrier2014}; see also the region $a \leq 5$\,AU of Figure~2 in \citealt{Jumper2013}).

\begin{table*}[t]

%\resizebox{0.75\textheight}{!}
%{\begin{minipage}{1.1\textwidth}

\centering
\caption{Brown dwarf companions in the Lick giant star survey sample. The masses are taken from \citet{Stock2018}, and the metallicities from \cite{Hekker2}.}
\label{table:Lick_BDs}

\begin{tabular}{lccrl}

\hline\hline  \noalign{\vskip 0.8mm}
Name & $m \sin i ~ [M_{\mathrm{Jup}}]$ & $M~[M_\odot]$ & [Fe/H] & Reference \\
\hline    \noalign{\vskip 0.8mm}
$\tau$\,Gem\,b & 22.1 & 2.47 & 0.14~ & \citet{Mitchell2013} \\
11\,Com\,b & 15.3 & 1.89 & $-0.24$~ & \citet{Liu2008} \\
$\nu$\,Oph\,b & 22.2 & 2.74 & 0.06~ & this paper \\
$\nu$\,Oph\,c & 24.7 & 2.74 & 0.06~ & this paper \\
\hline\hline
\end{tabular}
\end{table*}

The mass-dependence of the Brown Dwarf Desert seems to be borne out by comparing the numbers of brown dwarfs in RV surveys of main sequence stars and of giants. For example, the Keck-HIRES data on 1624 F to M dwarf stars tabulated by \citet{Butler2017} contain only two companions with $m \sin i$ in the brown dwarf range (HD\,16760\,b with $m \sin i = 13.1\,M_{\mathrm{Jup}}$ and $M = 0.78\,M_\odot$, and HD\,214823\,b with $m \sin i = 19.2\,M_{\mathrm{Jup}}$ and $M = 1.22\,M_\odot$). In contrast, our survey sample of 373 giant stars contains four brown dwarfs (see Tab.~\ref{table:Lick_BDs}). Taking these numbers at face value, the rate of incidence of brown dwarfs is nearly a factor of ten higher in the latter sample. The fact that all four companions have masses of roughly $20\,M_{\mathrm{Jup}}$ and orbit host stars of $\sim 2\,M_\odot$ or higher further support the hypothesis that these objects represent the high-mass end of the planetary mass function, which is shifted towards higher masses for more massive host stars.

One has to caution, however, that this is by no means a conclusive statistical analysis. In addition to the $\sin i$ ambiguity, such an analysis would have to address various selection effects that may affect the inferred companion rate: (1) For many surveys, there is no published information about the full underlying target sample. In those cases it is impossible to convert numbers of detections into occurrence rates, as additional unpublished companions may be contained in the full sample. The two examples above have been chosen because the full information of these surveys is available. (2) Essentially all large RV surveys have very uneven temporal sampling across the target stars, as the survey teams usually allocate additional observations to ``promising'' or ``interesting'' targets. This makes the companion detection thresholds rather non-uniform. (3) The task of establishing reliable detection thresholds is further complicated by varying levels of ``stellar noise'' due to activity and oscillations, whose amplitudes are strongly correlated with the stellar parameters, in particular for evolved stars \citep[e.g.,][]{Hekker}. (4) To analyze companion occurrence rates as a function of host star mass, these masses have to be determined in the first place. This is not a trivial task in the case of giant stars \citep{Stock2018}. (5) The occurrence rate of massive planets depends not only on mass, but also on metallicity \citep{Fischer2005, Reffert2015}. While neglecting this dependency may lead to erroneous conclusions, attempts to determine the occurrence rate of rare objects (such as brown dwarfs) as a function of both parameters suffer strongly from low-number statistics.

The resolution of these problems will likely have to wait for the release of the epoch astrometry data from the {\em Gaia} mission. A $10\,M_{\mathrm{Jup}}$ companion in a 1\,AU orbit around a $2\,M_\odot$ star at a distance of 100\,pc induces an astrometric motion of the host star with a 50\,$\mu$as amplitude, which should be readily detectable by {\em Gaia}. The mission will thus conduct a complete census of brown dwarf companions to thousands of A and F main sequence stars. For the time being, the double brown dwarf system $\nu$\,Oph provides additional insights into the link between brown dwarfs and high-mass planets, as discussed in the following section.

\subsection{The formation of brown dwarf companions}
\label{BDformation}

Brown dwarf companions can in principle form through three different mechanisms: (1) turbulent fragmentation of a molecular cloud \citep{Bate2012, Luhman2012}, i.e., like a binary star with very high mass ratio; (2) fragmentation of the disk during the formation of the primary \citep{Stamatellos2011, Kratter2016}; and (3) core accretion, i.e., like a Jovian planet. Models of these processes predict different properties of the companion population, but they could all contribute to a varying extent in different sections of the parameter space, which makes it difficult to assess their relative importance \citep{Marks2017}. It would thus be highly desirable to firmly identify the formation mechanism for individual well-characterized objects. The $\nu$\,Oph double-brown dwarf system is particularly suited to address this question, because its unusual properties place constraints on its formation pathway.

First we note that molecular cloud fragmentation into multiple ``star''-forming cores is a very unlikely formation process for the $\nu$\,Oph system. The finding that the two brown dwarfs are in a 6:1 MMR configuration is robust across virtually all of the allowed parameter space, for coplanar as well as mutually inclined orbits. No similar configuration is known for any multiple stellar system \citep{Tokovinin2018}, whereas orbital resonances are rather common in planetary systems, and their presence is best explained by resonance capture during convergent migration of the planets forming in the circumstellar disk \citep[e.g.,][]{Lee2002, Kley2012}. It is thus very likely that the two brown dwarfs formed in the disk of $\nu$\,Oph when it was a Herbig Ae star.

This leaves the question open whether disk fragmentation or core accretion is the more likely formation process. Unfortunately, we do not have access to the brown dwarfs' radii, and therefore there are no constraints on their composition, which could potentially discriminate between the two scenarios and provide additional information on the disk properties \citep{Guillot2014, Humphries2018}. There is no agreement in the literature about the importance of gravitational instabilities in circumstellar disks for planet and brown dwarf formation. While, e.g., \citet{Chabrier2014} dismiss this mechanism as a major contributor to the companion population around single stars, others argue that disk fragmentation is responsible for most of the companions with high masses and / or large orbital radii \citep{Stamatellos2009, Schlaufman2018, Vorobyov2018}. In fact, \citet{Dodson2009} investigated core accretion, scattering from the inner disk, and gravitational instability as potential formation scenarios; they conclude that only the last is a viable mechanism to form gas giants on stable orbits with semi-major axes $\gtrsim 35$\,AU. Simulations of self-gravitating disk fragments by \citet{Forgan2015} produced some systems that bear a remarkable resemblance to $\nu$\,Oph (see their Figure~2). This would seem to argue that $\nu$\,Oph\,b and $\nu$\,Oph\,c formed by disk fragmentation. One should caution, however, that these studies considered mostly host stars with $\sim 1\,M_\odot$. If the maximum mass of $10\,M_{\mathrm{Jup}}$ postulated by \citet{Schlaufman2018} for companions formed by core accretion scales with the host star mass, $\nu$\,Oph\,b and $\nu$\,Oph\,c may well fall below this limit.

\citet{Maldonado2017} investigate the dependence of brown dwarf formation on stellar metallicity and suggest that gravitational instability might be dominant at lower, and core accretion at higher values of the metallicity. Of the host stars to brown dwarfs in the Lick giant star survey sample (Tab.~\ref{table:Lick_BDs}), 11\,Com has a rather low metallicity (rank 78\,/\,366 of all stars in \citet{Hekker2}), whereas $\nu$\,Oph and $\tau$\,Gem have rather high metallicities (rank 318 and 354\,/\,366, respectively). With this small number of objects it is thus not possible to attribute the occurrence of brown dwarf companions among the Lick sample to either high or low host star metallicity.

There may be an interesting direct link between the $\nu$\,Oph system and planets in wide orbits found by direct imaging. For example, the HR\,8799 system also appears to have formed in a massive disk, and it has been suggested that its configuration represents a chain of multiple MMRs \citep{Fabrycky2010, Gozdziewski2014, Wang2018}. It thus appears plausible that both systems formed through fragmentation in the outer regions of the circumstellar disk \citep[$\gtrsim 40 \ldots 70$\,AU;][]{Kratter2010} and subsequent inward migration and resonance capture. However, in this scenario one would expect an even larger population of similar objects with larger companion masses \citep{Kratter2010, MurrayClay2010}, which appears not to be the case for either the directly imaged planets or the companions of giant stars found in RV surveys.

To settle the question about the origin of $\sim 20\,M_{\mathrm{Jup}}$ objects, better statistics and in particular more multiple systems are clearly needed. Since the number of giant stars accessible to radial-velocity observations with modest-size telescopes is far larger than the samples surveyed so far, additional examples may be found soon. At present, the system most similar to $\nu$\,Oph in the literature is BD\,+20\,2457, with two companions with masses $12.5\,M_{\mathrm{Jup}}$ and $21.4\,M_{\mathrm{Jup}}$ orbiting at 1.4\,AU and 2\,AU, respectively \citep{Niedzielski_b}. This system can probably not be stable if it is not in a MMR, but so far attempts to find any such stable configurations have failed, casting some doubts on the reality of the companions \citep{Horner2014, Trifonov2014}. It would thus appear highly desirable to collect more data on this star.

Another system that bears some close resemblance to $\nu$\,Oph is the 5:1 MMR system HD\,202206, which contains a brown dwarf ($ m \sin i = 16.6\,M_{\mathrm{Jup}}$) and a giant planet ($ m \sin i = 2.2\,M_{\mathrm{Jup}}$) orbiting a $1.0\,M_\odot$ star \citep{Correia2005, Couetdic2010}.
The HD\,202206 system can be explained by either formation of both brown dwarf and planet in a circumstellar disk, or formation of the star-brown dwarf binary and then formation of the planet in a circum-binary disk.

As a final remark we note that the true masses of $\nu$\,Oph\,b and $\nu$\,Oph\,c could be substantially larger than $20\,M_{\mathrm{Jup}}$, perhaps even above the hydrogen burning limit, as discussed in Sect.~\ref{Orbital fit}. This would not invalidate any of the arguments about the formation scenarios, but in that case core accretion does not appear plausible, leaving disk fragmentation as the only viable formation mechanism.

\section{Summary and conclusions}
\label{Summary}

In this paper we present results of an orbital analysis for the
$\nu$\,Oph system, which contains
an evolved K-giant star with $M = 2.7\,M_{\odot}$ and two brown dwarf
companions with minimum masses $m_1 \sin i = 22.2\,M_{\mathrm{Jup}}$ for the inner
companion, and $m_2 \sin i = 24.7\,M_{\mathrm{Jup}}$ for the outer.
It was previously known that this system is consistent with orbital periods close to 6:1 in ratio,
and thus we have performed a detailed study to decide if this system is indeed trapped in a mean motion resonance.

For our analysis we use 150 precise Lick Observatory Doppler measurements, which we obtained between 2000 and 2011.
We combine our RVs with an additional 44 precise RVs from the OAO Observatory, available in the literature.
The total 194 data points have a mean precision of $\sim 5$\,m\,s$^{-1}$, but we
quadratically added to the RV data uncertainties
an additional estimated stellar jitter velocity of 7.5\,m\,s$^{-1}$. Finally, we model the combined data with self-consistent dynamical fits,
which calculate $\nu$\,Oph's spectroscopic reflex motion
by taking into account the mutual gravitational interactions between the bodies in the system.

Our best coplanar edge-on model to the combined data
%has $\chi_{\nu}^2 = 1.042$,
%yielding orbital periods of $P_1 = 530.21 \pm 0.10$\,d,
%$P_2 = 3184.83 \pm 5.93$\,d, eccentricities of $e_1 = 0.124 \pm 0.003$ and $e_2 = 0.180 \pm 0.006$,
%and estimated minimum masses of $m_1 = 22.2\,M_{\mathrm{Jup}}$ and $m_2 = 24.7\,M_{\mathrm{Jup}}$.
%This model
suggests that both orbits are aligned with arguments of
periastron $\omega_1 = 9.9^\circ \pm 1.5^\circ$ and $\omega_2 = 8.3^\circ \pm 2.0^\circ$, respectively.
A long-term stability analysis reveals that the coplanar edge-on fit is stable for at least 1\,Gyr, and that the system is deeply trapped in a 6:1 MMR.
Our dynamical test reveals a very coherent orbital evolution, where the brown dwarf eccentricities librate with moderate amplitudes, while
the variations of the semimajor axes are negligible.
Close inspection of the long-term resonant state of the system shows that
all six resonance angles $\theta_{1}, \ldots, \theta_{6}$ and the secular $\Delta\omega$ are librating around 0$^\circ$.
We conclude that these results present strong evidence for the $\nu$\,Oph system being in a resonant configuration.

To verify the best edge-on coplanar model estimates for the $\nu$\,Oph system
we test a large number of synthetic RV data sets generated with a bootstrap technique.
We find that the achieved bootstrap distribution of the orbital elements and their
confidence levels are fully consistent with our best N-body edge-on fit and its error estimates.
In a final stability test we find that all bootstrap fits are stable for at least 10\,Myr, and that all are in resonance.

We have also carried out a large number of coplanar inclined dynamical fits in the same way as in the edge-on case.
The best inclined fit
%has $\chi_{\nu}^2 =1.037$, and
yields an inclination of $i = 16^\circ$, leading to much larger companions masses,
namely $m_1 = 82.0\,M_{\mathrm{Jup}}$ and $m_2 = 92.0\,M_{\mathrm{Jup}}$.
The remaining spectroscopic parameters for the best inclined fit, however,
are very similar to those from the coplanar edge-on model and
we find that both best-fit solutions mutually agree within the 1$\sigma$ uncertainties.
Since the improvement in $\chi^2$ compared to the edge-on fit is insignificant, we conclude that with the current combined RV data set it is impossible to
constrain the line of sight orbital inclination using a dynamical model.
%Individual test to the OAO and the Lick data showed that
Remarkably, all coplanar inclined fits down to $i = 5^\circ$ are also stable
and in 6:1 resonance. As expected, with increasing brown dwarf masses
the system becomes more dynamically active: the resonance angles still librate around 0$^\circ$, but with increasing libration amplitudes and frequencies.
In these fits the companions remain well separated and retains a clear 6:1 period ratio during the integrations.

Finally, we study the $\chi_{\nu}^2$ and stability of the system as a function of various orbital parameter combinations.
We construct high-density 2D coplanar edge-on grids with $P_2 / P_1$ vs.\ $P_1$,
$P_2$ vs.\ $e_2$ and $\omega_2$ vs.\ $e_2$ as fixed parameters, which were systematically varied on the grids,
while all other orbital parameters were allowed to vary freely.
We selected the parameter pairs in such a way that we could study the fit
properties out to at least a few $\sigma$ away from the best fit.
We find that all fits on these grids are stable for at least 10\,Myr.
The vast majority of these configurations are found to be in the 6:1 MMR. We repeat the 2D grid test for $i$ = 30$^\circ$, where the companion masses
are doubled, and find similar results.
%All investigated configurations were stable, and the fits within 3$\sigma$ (and above)
%are in a 6:1 MMR.
Since the two companions are well separated, increasing the companion masses has little influence on the long-term stability.

We also construct twelve grids with mutually inclined orbits by adopting different $\Delta\Omega$.
We conclude that moderate mutual inclinations in prograde orbits,
or small mutual inclinations in retrograde orbits are stable.
Except for a few isolated cases, mutual inclinations with $\Delta i$ between 60$^\circ$ and 150$^\circ$
lead to instability on very short time scales, most likely due to Kozai-Lidov effects.
We caution, however, that these stability tests were carried out only for 1\,Myr, and that longer integrations might reduce the sizes of the stable regions.

In summary, we conclude that the K giant star $\nu$\,Oph is orbited by two companions with minimum dynamical masses of $m_1 \sin i = 22.2\,M_{\mathrm{Jup}}$ and $m_2 \sin i = 24.7\,M_{\mathrm{Jup}}$, which are locked in a 6:1 MMR. This conclusion is robust also if large inclinations with respect to the line-of-sight or even mutually inclined orbits are considered. It is very likely that the two brown dwarf companions formed in the disk of $\nu$\,Oph when the system was young, but it is not possible at present to decide whether the mechanism was gravitational instability or core accretion.

\begin{acknowledgements}
We thank the staff of Lick Observatory for their excellent support over many years.
David Mitchell, Saskia Hekker, Simon Albrecht, Christian Schwab, Julian St\"{u}rmer and
all the graduate students that have spent many nights on Mt.\ Hamilton to collect spectra. Special thanks are
due to Geoff Marcy, Paul Butler, and Debra Fischer for permission to use their equipment
and software. T.T.\ and M.H.L.\ were supported in part by the Hong Kong RGC grant HKU 7024/13P.

\end{acknowledgements}

\bibliographystyle{aa}
\bibliography{Titos_biblio}

\begin{thebibliography}{86}
\expandafter\ifx\csname natexlab\endcsname\relax\def\natexlab#1{#1}\fi

\bibitem[{{Barban} {et~al.}(2004){Barban}, {De Ridder}, {Mazumdar}, {Carrier},
  {Eggenberger}, {De Ruyter}, {Vanautgaerden}, {Bouchy}, \& {Aerts}}]{Barban}
{Barban}, C., {De Ridder}, J., {Mazumdar}, A., {et~al.} 2004, in ESA Special
  Publication, Vol. 559, SOHO 14 Helio- and Asteroseismology: Towards a Golden
  Future, ed. D.~{Danesy}, 113

\bibitem[{{Bate}(2012)}]{Bate2012}
{Bate}, M.~R. 2012, \mnras, 419, 3115

\bibitem[{{Boss} {et~al.}(2007){Boss}, {Butler}, {Hubbard}, {Ianna},
  {K{\"u}rster}, {Lissauer}, {Mayor}, {Meech}, {Mignard}, {Penny},
  {Quirrenbach}, {Tarter}, \& {Vidal-Madjar}}]{IAU2007}
{Boss}, A.~P., {Butler}, R.~P., {Hubbard}, W.~B., {et~al.} 2007, Transactions
  of the International Astronomical Union, Series A, 26, 183

\bibitem[{{Bryden} {et~al.}(2000){Bryden}, {R{\'o}{\.z}yczka}, {Lin}, \&
  {Bodenheimer}}]{Bryden2000}
{Bryden}, G., {R{\'o}{\.z}yczka}, M., {Lin}, D.~N.~C., \& {Bodenheimer}, P.
  2000, \apj, 540, 1091

\bibitem[{{Butler} {et~al.}(1997){Butler}, {Marcy}, {Williams}, {Hauser}, \&
  {Shirts}}]{Butler1997}
{Butler}, R.~P., {Marcy}, G.~W., {Williams}, E., {Hauser}, H., \& {Shirts}, P.
  1997, \apjl, 474, L115

\bibitem[{{Butler} {et~al.}(1996){Butler}, {Marcy}, {Williams}, {McCarthy},
  {Dosanjh}, \& {Vogt}}]{Butler}
{Butler}, R.~P., {Marcy}, G.~W., {Williams}, E., {et~al.} 1996, \pasp, 108, 500

\bibitem[{{Butler} {et~al.}(2017){Butler}, {Vogt}, {Laughlin}, {Burt},
  {Rivera}, {Tuomi}, {Teske}, {Arriagada}, {Diaz}, {Holden}, \&
  {Keiser}}]{Butler2017}
{Butler}, R.~P., {Vogt}, S.~S., {Laughlin}, G., {et~al.} 2017, \aj, 153, 208

\bibitem[{{Chabrier} {et~al.}(2014){Chabrier}, {Johansen}, {Janson}, \&
  {Rafikov}}]{Chabrier2014}
{Chabrier}, G., {Johansen}, A., {Janson}, M., \& {Rafikov}, R. 2014, Protostars
  and Planets VI, 619

\bibitem[{{Correia} {et~al.}(2009){Correia}, {Udry}, {Mayor}, {Benz},
  {Bertaux}, {Bouchy}, {Laskar}, {Lovis}, {Mordasini}, {Pepe}, \&
  {Queloz}}]{Correia}
{Correia}, A.~C.~M., {Udry}, S., {Mayor}, M., {et~al.} 2009, \aap, 496, 521

\bibitem[{{Correia} {et~al.}(2005){Correia}, {Udry}, {Mayor}, {Laskar}, {Naef},
  {Pepe}, {Queloz}, \& {Santos}}]{Correia2005}
{Correia}, A.~C.~M., {Udry}, S., {Mayor}, M., {et~al.} 2005, \aap, 440, 751

\bibitem[{{Couetdic} {et~al.}(2010){Couetdic}, {Laskar}, {Correia}, {Mayor}, \&
  {Udry}}]{Couetdic2010}
{Couetdic}, J., {Laskar}, J., {Correia}, A.~C.~M., {Mayor}, M., \& {Udry}, S.
  2010, \aap, 519, A10

\bibitem[{{De Ridder} {et~al.}(2006){De Ridder}, {Barban}, {Carrier},
  {Mazumdar}, {Eggenberger}, {Aerts}, {Deruyter}, \&
  {Vanautgaerden}}]{De_Ridder}
{De Ridder}, J., {Barban}, C., {Carrier}, F., {et~al.} 2006, \aap, 448, 689

\bibitem[{{Deleuil} {et~al.}(2008){Deleuil}, {Deeg}, {Alonso}, {Bouchy},
  {Rouan}, {Auvergne}, {Baglin}, {Aigrain}, {Almenara}, {Barbieri}, {Barge},
  {Bruntt}, {Bord{\'e}}, {Collier Cameron}, {Csizmadia}, {de La Reza},
  {Dvorak}, {Erikson}, {Fridlund}, {Gandolfi}, {Gillon}, {Guenther}, {Guillot},
  {Hatzes}, {H{\'e}brard}, {Jorda}, {Lammer}, {L{\'e}ger}, {Llebaria},
  {Loeillet}, {Mayor}, {Mazeh}, {Moutou}, {Ollivier}, {P{\"a}tzold}, {Pont},
  {Queloz}, {Rauer}, {Schneider}, {Shporer}, {Wuchterl}, \&
  {Zucker}}]{Deleuil2008}
{Deleuil}, M., {Deeg}, H.~J., {Alonso}, R., {et~al.} 2008, \aap, 491, 889

\bibitem[{{Desort} {et~al.}(2008){Desort}, {Lagrange}, {Galland}, {Beust},
  {Udry}, {Mayor}, \& {Lo Curto}}]{Desort2008}
{Desort}, M., {Lagrange}, A.-M., {Galland}, F., {et~al.} 2008, \aap, 491, 883

\bibitem[{{Dodson-Robinson} {et~al.}(2009){Dodson-Robinson}, {Veras}, {Ford},
  \& {Beichman}}]{Dodson2009}
{Dodson-Robinson}, S.~E., {Veras}, D., {Ford}, E.~B., \& {Beichman}, C.~A.
  2009, \apj, 707, 79

\bibitem[{{Duncan} {et~al.}(1998){Duncan}, {Levison}, \& {Lee}}]{Duncan1998}
{Duncan}, M.~J., {Levison}, H.~F., \& {Lee}, M.~H. 1998, \aj, 116, 2067

\bibitem[{{ESA}(1997)}]{ESA1997}
{ESA}. 1997, ESA Special Publication, Vol. 1200, {The HIPPARCOS and TYCHO
  catalogues. Astrometric and photometric star catalogues derived from the ESA
  HIPPARCOS Space Astrometry Mission}

\bibitem[{{Fabrycky} \& {Murray-Clay}(2010)}]{Fabrycky2010}
{Fabrycky}, D.~C. \& {Murray-Clay}, R.~A. 2010, \apj, 710, 1408

\bibitem[{{Fischer} \& {Valenti}(2005)}]{Fischer2005}
{Fischer}, D.~A. \& {Valenti}, J. 2005, \apj, 622, 1102

\bibitem[{{Fischer} {et~al.}(2007){Fischer}, {Vogt}, {Marcy}, {Butler}, {Sato},
  {Henry}, {Robinson}, {Laughlin}, {Ida}, {Toyota}, {Omiya}, {Driscoll},
  {Takeda}, {Wright}, \& {Johnson}}]{Fischer2007}
{Fischer}, D.~A., {Vogt}, S.~S., {Marcy}, G.~W., {et~al.} 2007, \apj, 669, 1336

\bibitem[{{Ford}(2005)}]{Ford}
{Ford}, E.~B. 2005, \aj, 129, 1706

\bibitem[{{Forgan} {et~al.}(2015){Forgan}, {Parker}, \& {Rice}}]{Forgan2015}
{Forgan}, D., {Parker}, R.~J., \& {Rice}, K. 2015, \mnras, 447, 836

\bibitem[{{Frink} {et~al.}(2002){Frink}, {Mitchell}, {Quirrenbach}, {Fischer},
  {Marcy}, \& {Butler}}]{Frink2}
{Frink}, S., {Mitchell}, D.~S., {Quirrenbach}, A., {et~al.} 2002, \apj, 576,
  478

\bibitem[{{Frink} {et~al.}(2001){Frink}, {Quirrenbach}, {Fischer}, {R{\"o}ser},
  \& {Schilbach}}]{Frink}
{Frink}, S., {Quirrenbach}, A., {Fischer}, D., {R{\"o}ser}, S., \& {Schilbach},
  E. 2001, \pasp, 113, 173

\bibitem[{{Girardi} {et~al.}(2000){Girardi}, {Bressan}, {Bertelli}, \&
  {Chiosi}}]{Girardi2000}
{Girardi}, L., {Bressan}, A., {Bertelli}, G., \& {Chiosi}, C. 2000, \aaps, 141,
  371

\bibitem[{{Go{\'z}dziewski} \& {Migaszewski}(2014)}]{Gozdziewski2014}
{Go{\'z}dziewski}, K. \& {Migaszewski}, C. 2014, \mnras, 440, 3140

\bibitem[{{Grether} \& {Lineweaver}(2006)}]{Grether2006}
{Grether}, D. \& {Lineweaver}, C.~H. 2006, \apj, 640, 1051

\bibitem[{{Guillot} {et~al.}(2014){Guillot}, {Lin}, {Morel}, {Havel}, \&
  {Parmentier}}]{Guillot2014}
{Guillot}, T., {Lin}, D.~N.~C., {Morel}, P., {Havel}, M., \& {Parmentier}, V.
  2014, in EAS Publications Series, Vol.~65, The Ages of Stars, 327

\bibitem[{{Halbwachs} {et~al.}(2000){Halbwachs}, {Arenou}, {Mayor}, {Udry}, \&
  {Queloz}}]{Halbwachs2000}
{Halbwachs}, J.~L., {Arenou}, F., {Mayor}, M., {Udry}, S., \& {Queloz}, D.
  2000, \aap, 355, 581

\bibitem[{{Hekker} \& {Mel{\'e}ndez}(2007)}]{Hekker2}
{Hekker}, S. \& {Mel{\'e}ndez}, J. 2007, \aap, 475, 1003

\bibitem[{{Hekker} {et~al.}(2006){Hekker}, {Reffert}, {Quirrenbach},
  {Mitchell}, {Fischer}, {Marcy}, \& {Butler}}]{Hekker}
{Hekker}, S., {Reffert}, S., {Quirrenbach}, A., {et~al.} 2006, \aap, 454, 943

\bibitem[{{Horner} {et~al.}(2014){Horner}, {Wittenmyer}, {Hinse}, \&
  {Marshall}}]{Horner2014}
{Horner}, J., {Wittenmyer}, R.~A., {Hinse}, T.~C., \& {Marshall}, J.~P. 2014,
  \mnras, 439, 1176

\bibitem[{{Humphries} \& {Nayakshin}(2018)}]{Humphries2018}
{Humphries}, R.~J. \& {Nayakshin}, S. 2018, \mnras, 477, 593

\bibitem[{{Izumiura}(1999)}]{Izumiura1999}
{Izumiura}, H. 1999, Publications of the Yunnan Observatory, 77

\bibitem[{{Johnson} {et~al.}(2006){Johnson}, {Marcy}, {Fischer}, {Henry},
  {Wright}, {Isaacson}, \& {McCarthy}}]{Johnson2006}
{Johnson}, J.~A., {Marcy}, G.~W., {Fischer}, D.~A., {et~al.} 2006, \apj, 652,
  1724

\bibitem[{{Jumper} \& {Fisher}(2013)}]{Jumper2013}
{Jumper}, P.~H. \& {Fisher}, R.~T. 2013, \apj, 769, 9

\bibitem[{{Kjeldsen} \& {Bedding}(1995)}]{Kjeldsen1995}
{Kjeldsen}, H. \& {Bedding}, T.~R. 1995, \aap, 293, 87

\bibitem[{{Kley}(2000)}]{Kley2000}
{Kley}, W. 2000, \mnras, 313, L47

\bibitem[{{Kley} \& {Nelson}(2012)}]{Kley2012}
{Kley}, W. \& {Nelson}, R.~P. 2012, \araa, 50, 211

\bibitem[{{Kratter} \& {Lodato}(2016)}]{Kratter2016}
{Kratter}, K. \& {Lodato}, G. 2016, \araa, 54, 271

\bibitem[{{Kratter} {et~al.}(2010){Kratter}, {Murray-Clay}, \&
  {Youdin}}]{Kratter2010}
{Kratter}, K.~M., {Murray-Clay}, R.~A., \& {Youdin}, A.~N. 2010, \apj, 710,
  1375

\bibitem[{{Lee} {et~al.}(2006){Lee}, {Butler}, {Fischer}, {Marcy}, \&
  {Vogt}}]{Lee2006}
{Lee}, M.~H., {Butler}, R.~P., {Fischer}, D.~A., {Marcy}, G.~W., \& {Vogt},
  S.~S. 2006, \apj, 641, 1178

\bibitem[{{Lee} \& {Peale}(2002)}]{Lee2002}
{Lee}, M.~H. \& {Peale}, S.~J. 2002, \apj, 567, 596

\bibitem[{{Lee} \& {Peale}(2003)}]{Lee2003}
{Lee}, M.~H. \& {Peale}, S.~J. 2003, \apj, 592, 1201

\bibitem[{{Liu} {et~al.}(2008){Liu}, {Sato}, {Zhao}, {Noguchi}, {Wang},
  {Kambe}, {Ando}, {Izumiura}, {Chen}, {Okada}, {Toyota}, {Omiya}, {Masuda},
  {Takeda}, {Murata}, {Itoh}, {Yoshida}, {Kokubo}, \& {Ida}}]{Liu2008}
{Liu}, Y.-J., {Sato}, B., {Zhao}, G., {et~al.} 2008, \apj, 672, 553

\bibitem[{{Luhman}(2012)}]{Luhman2012}
{Luhman}, K.~L. 2012, \araa, 50, 65

\bibitem[{{Maldonado} \& {Villaver}(2017)}]{Maldonado2017}
{Maldonado}, J. \& {Villaver}, E. 2017, \aap, 602, A38

\bibitem[{{Marcy} \& {Butler}(2000)}]{Marcy2000}
{Marcy}, G.~W. \& {Butler}, R.~P. 2000, \pasp, 112, 137

\bibitem[{{Marcy} {et~al.}(2001){Marcy}, {Butler}, {Fischer}, {Vogt},
  {Lissauer}, \& {Rivera}}]{Marcy}
{Marcy}, G.~W., {Butler}, R.~P., {Fischer}, D., {et~al.} 2001, \apj, 556, 296

\bibitem[{{Marks} {et~al.}(2017){Marks}, {Mart{\'{\i}}n}, {B{\'e}jar},
  {Lodieu}, {Kroupa}, {Manjavacas}, {Thies}, {Rebolo L{\'o}pez}, \&
  {Velasco}}]{Marks2017}
{Marks}, M., {Mart{\'{\i}}n}, E.~L., {B{\'e}jar}, V.~J.~S., {et~al.} 2017,
  \aap, 605, A11

\bibitem[{{Mayor} \& {Queloz}(1995)}]{Mayor1995}
{Mayor}, M. \& {Queloz}, D. 1995, \nat, 378, 355

\bibitem[{{Mitchell} {et~al.}(2003){Mitchell}, {Frink}, {Quirrenbach},
  {Fischer}, {Marcy}, \& {Butler}}]{Mitchell2003}
{Mitchell}, D.~S., {Frink}, S., {Quirrenbach}, A., {et~al.} 2003, in BAAS,
  Vol.~35, American Astronomical Society Meeting Abstracts, 1234

\bibitem[{{Mitchell} {et~al.}(2013){Mitchell}, {Reffert}, {Trifonov},
  {Quirrenbach}, \& {Fischer}}]{Mitchell2013}
{Mitchell}, D.~S., {Reffert}, S., {Trifonov}, T., {Quirrenbach}, A., \&
  {Fischer}, D.~A. 2013, \aap, 555, A87

\bibitem[{{Murray-Clay}(2010)}]{MurrayClay2010}
{Murray-Clay}, R.~A. 2010, in Astronomical Society of the Pacific Conference
  Series, Vol. 432, New Horizons in Astronomy: Frank N. Bash Symposium 2009,
  ed. L.~M. {Stanford}, J.~D. {Green}, L.~{Hao}, \& Y.~{Mao}, 98

\bibitem[{{Nidever} {et~al.}(2002){Nidever}, {Marcy}, {Butler}, {Fischer}, \&
  {Vogt}}]{Nidever2002}
{Nidever}, D.~L., {Marcy}, G.~W., {Butler}, R.~P., {Fischer}, D.~A., \& {Vogt},
  S.~S. 2002, The Astrophysical Journal Supplement Series, 141, 503

\bibitem[{{Niedzielski} {et~al.}(2007){Niedzielski}, {Konacki}, {Wolszczan},
  {Nowak}, {Maciejewski}, {Gelino}, {Shao}, {Shetrone}, \&
  {Ramsey}}]{Niedzielski2007}
{Niedzielski}, A., {Konacki}, M., {Wolszczan}, A., {et~al.} 2007, \apj, 669,
  1354

\bibitem[{{Niedzielski} {et~al.}(2009){Niedzielski}, {Nowak}, {Adam{\'o}w}, \&
  {Wolszczan}}]{Niedzielski_b}
{Niedzielski}, A., {Nowak}, G., {Adam{\'o}w}, M., \& {Wolszczan}, A. 2009,
  \apj, 707, 768

\bibitem[{{Ortiz} {et~al.}(2016){Ortiz}, {Reffert}, {Trifonov}, {Quirrenbach},
  {Mitchell}, {Nowak}, {Buenzli}, {Zimmerman}, {Bonnefoy}, {Skemer},
  {Defr{\`e}re}, {Lee}, {Fischer}, \& {Hinz}}]{Ortiz2016}
{Ortiz}, M., {Reffert}, S., {Trifonov}, T., {et~al.} 2016, \aap, 595, A55

\bibitem[{{Pascucci} {et~al.}(2016){Pascucci}, {Testi}, {Herczeg}, {Long},
  {Manara}, {Hendler}, {Mulders}, {Krijt}, {Ciesla}, {Henning}, {Mohanty},
  {Drabek-Maunder}, {Apai}, {Sz{\H u}cs}, {Sacco}, \&
  {Olofsson}}]{Pascucci2016}
{Pascucci}, I., {Testi}, L., {Herczeg}, G.~J., {et~al.} 2016, \apj, 831, 125

\bibitem[{{Patel} {et~al.}(2007){Patel}, {Vogt}, {Marcy}, {Johnson}, {Fischer},
  {Wright}, \& {Butler}}]{Patel2007}
{Patel}, S.~G., {Vogt}, S.~S., {Marcy}, G.~W., {et~al.} 2007, \apj, 665, 744

\bibitem[{{Press} {et~al.}(1992){Press}, {Teukolsky}, {Vetterling}, \&
  {Flannery}}]{Press}
{Press}, W.~H., {Teukolsky}, S.~A., {Vetterling}, W.~T., \& {Flannery}, B.~P.
  1992, {Numerical recipes in FORTRAN. The art of scientific computing}
  (Cambridge University Press, 2nd ed.)

\bibitem[{{Quirrenbach} {et~al.}(2011){Quirrenbach}, {Reffert}, \&
  {Bergmann}}]{Quirrenbach}
{Quirrenbach}, A., {Reffert}, S., \& {Bergmann}, C. 2011, in American Institute
  of Physics Conference Series, Vol. 1331, Planetary Systems Beyond the Main
  Sequence, ed. S.~{Schuh}, H.~{Drechsel}, \& U.~{Heber}, 102

\bibitem[{{Reffert} {et~al.}(2015){Reffert}, {Bergmann}, {Quirrenbach},
  {Trifonov}, \& {K{\"u}nstler}}]{Reffert2015}
{Reffert}, S., {Bergmann}, C., {Quirrenbach}, A., {Trifonov}, T., \&
  {K{\"u}nstler}, A. 2015, \aap, 574, A116

\bibitem[{{Reffert} \& {Quirrenbach}(2011)}]{Reffert2011}
{Reffert}, S. \& {Quirrenbach}, A. 2011, \aap, 527, A140

\bibitem[{{Reffert} {et~al.}(2006){Reffert}, {Quirrenbach}, {Mitchell},
  {Albrecht}, {Hekker}, {Fischer}, {Marcy}, \& {Butler}}]{Reffert}
{Reffert}, S., {Quirrenbach}, A., {Mitchell}, D.~S., {et~al.} 2006, \apj, 652,
  661

\bibitem[{{Sahlmann} {et~al.}(2011){Sahlmann}, {S{\'e}gransan}, {Queloz},
  {Udry}, {Santos}, {Marmier}, {Mayor}, {Naef}, {Pepe}, \&
  {Zucker}}]{Sahlmann2011}
{Sahlmann}, J., {S{\'e}gransan}, D., {Queloz}, D., {et~al.} 2011, \aap, 525,
  A95

\bibitem[{{Sato} {et~al.}(2003){Sato}, {Ando}, {Kambe}, {Takeda}, {Izumiura},
  {Masuda}, {Watanabe}, {Noguchi}, {Wada}, {Okada}, {Koyano}, {Maehara},
  {Norimoto}, {Okada}, {Shimizu}, {Uraguchi}, {Yanagisawa}, \&
  {Yoshida}}]{Sato2003}
{Sato}, B., {Ando}, H., {Kambe}, E., {et~al.} 2003, \apjl, 597, L157

\bibitem[{{Sato} {et~al.}(2012){Sato}, {Omiya}, {Harakawa}, {Izumiura},
  {Kambe}, {Takeda}, {Yoshida}, {Itoh}, {Ando}, {Kokubo}, \& {Ida}}]{Sato2012}
{Sato}, B., {Omiya}, M., {Harakawa}, H., {et~al.} 2012, \pasj, 64, 135

\bibitem[{{Schlaufman}(2018)}]{Schlaufman2018}
{Schlaufman}, K.~C. 2018, \apj, 853, 37

\bibitem[{{Setiawan} {et~al.}(2003){Setiawan}, {Hatzes}, {von der L{\"u}he},
  {Pasquini}, {Naef}, {da Silva}, {Udry}, {Queloz}, \&
  {Girardi}}]{Setiawan2003}
{Setiawan}, J., {Hatzes}, A.~P., {von der L{\"u}he}, O., {et~al.} 2003, \aap,
  398, L19

\bibitem[{{Sozzetti} \& {Desidera}(2010)}]{Sozzetti2010}
{Sozzetti}, A. \& {Desidera}, S. 2010, \aap, 509, A103

\bibitem[{{Stamatellos} \& {Whitworth}(2011)}]{Stamatellos2011}
{Stamatellos}, D. \& {Whitworth}, A. 2011, in European Physical Journal Web of
  Conferences, Vol.~16, European Physical Journal Web of Conferences, 05001

\bibitem[{{Stamatellos} \& {Whitworth}(2009)}]{Stamatellos2009}
{Stamatellos}, D. \& {Whitworth}, A.~P. 2009, \mnras, 392, 413

\bibitem[{{Stock} {et~al.}(2018){Stock}, {Reffert}, \&
  {Quirrenbach}}]{Stock2018}
{Stock}, S., {Reffert}, S., \& {Quirrenbach}, A. 2018, \aap, 616, A33

\bibitem[{{Tan} {et~al.}(2013){Tan}, {Payne}, {Lee}, {Ford}, {Howard},
  {Johnson}, {Marcy}, \& {Wright}}]{Tan2013}
{Tan}, X., {Payne}, M.~J., {Lee}, M.~H., {et~al.} 2013, \apj, 777, 101

\bibitem[{{Tokovinin}(2018)}]{Tokovinin2018}
{Tokovinin}, A. 2018, \apjs, 235, 6

\bibitem[{{Trifonov} {et~al.}(2014){Trifonov}, {Reffert}, {Tan}, {Lee}, \&
  {Quirrenbach}}]{Trifonov2014}
{Trifonov}, T., {Reffert}, S., {Tan}, X., {Lee}, M.~H., \& {Quirrenbach}, A.
  2014, \aap, 568, A64

\bibitem[{{Trifonov} {et~al.}(2015){Trifonov}, {Reffert}, {Zechmeister},
  {Reiners}, \& {Quirrenbach}}]{Trifonov2015}
{Trifonov}, T., {Reffert}, S., {Zechmeister}, M., {Reiners}, A., \&
  {Quirrenbach}, A. 2015, \aap, 582, A54

\bibitem[{{Valenti} {et~al.}(1995){Valenti}, {Butler}, \& {Marcy}}]{Valenti}
{Valenti}, J.~A., {Butler}, R.~P., \& {Marcy}, G.~W. 1995, \pasp, 107, 966

\bibitem[{{van Leeuwen}(2007)}]{Leeuwen}
{van Leeuwen}, F. 2007, \aap, 474, 653

\bibitem[{{Vogt}(1987)}]{Vogt1987}
{Vogt}, S.~S. 1987, \pasp, 99, 1214

\bibitem[{{Vorobyov} \& {Elbakyan}(2018)}]{Vorobyov2018}
{Vorobyov}, E.~I. \& {Elbakyan}, V.~G. 2018, \aap, 618, A7

\bibitem[{{Wang} {et~al.}(2018){Wang}, {Graham}, {Dawson}, {Fabrycky}, {De
  Rosa}, {Pueyo}, {Konopacky}, {Macintosh}, {Marois}, {Chiang}, {Ammons},
  {Arriaga}, {Bailey}, {Barman}, {Bulger}, {Chilcote}, {Cotten}, {Doyon},
  {Duch{\^e}ne}, {Esposito}, {Fitzgerald}, {Follette}, {Gerard}, {Goodsell},
  {Greenbaum}, {Hibon}, {Hung}, {Ingraham}, {Kalas}, {Larkin}, {Maire},
  {Marchis}, {Marley}, {Metchev}, {Millar-Blanchaer}, {Nielsen}, {Oppenheimer},
  {Palmer}, {Patience}, {Perrin}, {Poyneer}, {Rajan}, {Rameau},
  {Rantakyr{\"o}}, {Ruffio}, {Savransky}, {Schneider}, {Sivaramakrishnan},
  {Song}, {Soummer}, {Thomas}, {Wallace}, {Ward-Duong}, {Wiktorowicz}, \&
  {Wolff}}]{Wang2018}
{Wang}, J.~J., {Graham}, J.~R., {Dawson}, R., {et~al.} 2018, \aj, 156, 192

\bibitem[{{Wilson} {et~al.}(2016){Wilson}, {H{\'e}brard}, {Santos}, {Sahlmann},
  {Montagnier}, {Astudillo-Defru}, {Boisse}, {Bouchy}, {Rey}, {Arnold},
  {Bonfils}, {Bourrier}, {Courcol}, {Deleuil}, {Delfosse}, {D{\'{\i}}az},
  {Ehrenreich}, {Forveille}, {Moutou}, {Pepe}, {Santerne}, {S{\'e}gransan}, \&
  {Udry}}]{Wilson2016}
{Wilson}, P.~A., {H{\'e}brard}, G., {Santos}, N.~C., {et~al.} 2016, \aap, 588,
  A144

\bibitem[{{Wright}(2005)}]{Wright2005}
{Wright}, J.~T. 2005, \pasp, 117, 657

\bibitem[{{Zechmeister} {et~al.}(2008){Zechmeister}, {Reffert}, {Hatzes},
  {Endl}, \& {Quirrenbach}}]{Zechmeister2008}
{Zechmeister}, M., {Reffert}, S., {Hatzes}, A.~P., {Endl}, M., \&
  {Quirrenbach}, A. 2008, \aap, 491, 531

\end{thebibliography}

\begin{appendix} %First online appendix

 \setcounter{table}{0}
\renewcommand{\thetable}{A\arabic{table}}

%------------------------------------------------------------------------------------------------------------

\begin{table*}{}%[hb]

\centering
\caption{Measured velocities for $\nu$\,Oph and the derived errors.}
\label{table:rvlick}
\resizebox{0.63\textheight}{!}{\begin{minipage}{\textwidth}

 \begin{tabular}{ crc  p{0.01mm} crc p{0.01mm} crc }
\cline{1-3}\cline{1-3}\cline{5-7}\cline{5-7}\cline{9-11}\cline{9-11}
\noalign{\vskip 0.5mm}
\cline{1-3}\cline{1-3}\cline{5-7}\cline{5-7}\cline{9-11}\cline{9-11}
\noalign{\vskip 0.5mm}

JD & RV [m\,s$^{-1}$] & $\sigma_{RV}$ [m\,s$^{-1}$] & &JD & RV [m\,s$^{-1}$] & $\sigma_{RV}$  [m\,s$^{-1}$] && JD & RV [m\,s$^{-1}$] & $\sigma_{RV}$  [m\,s$^{-1}$]\\
\cline{1-3}\cline{5-7}\cline{9-11}
\noalign{\vskip 0.5mm}

  2451853.595 &  $-$285.8 &  5.3 &&        2452495.757 &  237.5 &  6.1    &&   2453854.954 &  $-$341.6 &  4.8 \\
  2451854.601 &  $-$281.0 &  5.9 &&        2452496.825 &  244.2 &  5.7	  &&   2453911.889 &  $-$349.2 &  6.2 \\
  2451856.598 &  $-$273.8 &  5.5 &&        2452505.795 &  283.8 &  5.6	  &&   2453915.853 &  $-$345.3 &  4.5 \\
  2452012.028 &     185.7 &  5.7 &&        2452517.749 &  300.4 &  6.4	  &&   2453931.799 &  $-$342.7 &  5.1 \\
  2452014.001 &     194.5 &  6.2 &&        2452519.738 &  299.7 &  5.1	  &&   2453936.734 &  $-$357.1 &  8.0 \\
  2452044.920 &     205.2 &  4.9 &&        2452529.758 &  334.9 &  6.5	  &&   2453967.789 &  $-$273.8 &  4.7 \\
  2452045.972 &     203.7 &  5.4 &&        2452530.675 &  346.3 &  4.3	  &&   2453981.698 &  $-$244.2 &  4.6 \\
  2452046.949 &     198.4 &  5.4 &&        2452531.737 &  345.8 &  6.1	  &&   2454182.023 &     136.6 &  4.8 \\
  2452048.896 &     196.4 &  5.8 &&        2452532.720 &  351.8 &  5.8	  &&   2454226.985 &   $-$23.1 &  4.9 \\
  2452079.895 &     140.0 &  5.4 &&        2452533.669 &  340.4 &  5.7	  &&   2454254.845 &  $-$158.0 &  4.4 \\
  2452080.876 &     119.0 &  5.2 &&        2452541.698 &  356.8 &  5.4	  &&   2454265.881 &  $-$197.1 &  4.6 \\
  2452081.891 &     116.2 &  5.4 &&        2452542.697 &  377.1 &  6.1	  &&   2454297.779 &  $-$288.1 &  4.7 \\
  2452082.900 &     111.0 &  5.1 &&        2452543.678 &  363.7 &  5.9	  &&   2454314.776 &  $-$322.6 &  5.1 \\
  2452083.801 &     116.9 &  4.8 &&        2452560.663 &  348.4 &  6.1	  &&   2454344.713 &  $-$396.3 &  4.8 \\
  2452098.860 &      58.3 &  5.1 &&        2452571.631 &  352.0 &  5.8	  &&   2454391.630 &  $-$433.6 &  8.8 \\
  2452100.846 &      56.6 &  4.8 &&        2452572.610 &  359.7 &  5.7	  &&   2454421.582 &  $-$439.4 &  5.6 \\
  2452106.874 &      31.9 &  5.9 &&        2452573.640 &  346.3 &  7.0	  &&   2454507.096 &  $-$307.3 &  5.3 \\
  2452109.844 &      27.6 &  5.0 &&        2452590.584 &  350.6 &  6.8	  &&   2454583.884 &   $-$84.4 &  5.4 \\
  2452124.806 &   $-$24.1 &  4.8 &&        2452707.065 & $-$5.8 &  4.7	  &&   2454600.885 &    $-$7.1 &  5.1 \\
  2452125.794 &   $-$22.8 &  4.9 &&        2452718.044 & $-$37.0 &  5.0	  &&   2454645.846 &     107.2 &  5.4 \\
  2452156.675 &  $-$134.4 &  4.5 &&        2452720.986 & $-$37.2 &  4.8	  &&   2454667.829 &     130.6 &  5.6 \\
  2452163.736 &  $-$150.1 &  5.2 &&        2452765.953 &$-$102.7 &  6.5	  &&   2454683.737 &     137.9 &  4.8 \\
  2452165.692 &  $-$166.0 &  5.5 &&        2452800.885 &$-$135.8 &  5.6	  &&   2454711.705 &	  74.8 &  4.8 \\
  2452166.657 &  $-$168.3 &  4.6 &&        2452803.894 &$-$122.1 &  5.3	  &&   2454756.639 &   $-$66.5 &  4.5 \\
  2452175.668 &  $-$183.9 &  4.7 &&        2452837.808 &$-$111.2 &  6.7	  &&   2454911.981 &  $-$401.1 &  7.8 \\
  2452192.623 &  $-$227.4 &  4.4 &&        2452861.771 &$-$79.5 &  4.4	  &&   2454979.970 &  $-$374.6 &  7.3 \\
  2452194.646 &  $-$236.1 &  5.7 &&        2452862.770 &$-$76.9 &  5.4	  &&   2455026.775 &  $-$300.7 &  7.2 \\
  2452205.607 &  $-$244.0 &  5.4 &&        2452863.802 & $-$71.7 & 5.5	  &&   2455063.780 &  $-$200.5 &  6.4 \\
  2452206.610 &  $-$240.9 &  5.7 &&        2452879.776 & $-$57.7 & 5.0	  &&   2455098.699 &   $-$77.1 &  6.7 \\
  2452207.592 &  $-$249.2 &  5.8 &&        2452898.710 & $-$7.7 &  5.2	  &&   2455121.628 &    $-$5.2 &  7.3 \\
  2452222.586 &  $-$267.9 &  5.3 &&        2452900.695 & $-$12.2 &  5.7	  &&   2455242.086 &     182.1 &  9.4 \\
  2452308.077 &  $-$302.5 &  5.8 &&        2452932.634 & 74.0 &  5.4	  &&   2455278.014 &	  70.4 &  7.4 \\
  2452337.079 &  $-$250.5 &  5.7 &&        2452934.630 & 77.0 &  4.7	  &&   2455303.005 &   $-$11.1 &  5.4 \\
  2452362.961 &  $-$205.4 &  5.5 &&        2453093.021 & 482.0 &  6.4	  &&   2455328.964 &  $-$106.4 &  5.9 \\
  2452363.987 &  $-$197.3 &  5.8 &&        2453168.890 & 273.0 &  5.0	  &&   2455329.956 &  $-$114.5 &  5.6 \\
  2452383.993 &  $-$149.4 &  6.0 &&        2453232.768 & 21.8 &  4.5	  &&   2455362.853 &  $-$187.4 &  6.5 \\
  2452384.979 &  $-$132.9 &  6.0 &&        2453265.682 & $-$49.8 &  4.6	  &&   2455420.785 &  $-$304.7 &  5.8 \\
  2452394.967 &  $-$113.5 &  6.1 &&        2453267.661 & $-$70.2 &  4.6	  &&   2455450.677 &  $-$312.3 &  6.3 \\
  2452412.826 &   $-$66.1 &  6.3 &&        2453286.637 & $-$96.5 &  4.5	  &&   2455589.097 &   $-$70.8 &  4.9 \\
  2452423.900 &    $-$8.7 &  6.2 &&        2453291.619 &$-$111.3 &  4.1	  &&   2455591.096 &   $-$70.5 &  5.9 \\
  2452425.901 &       2.9 &  6.1 &&        2453293.624 &$-$111.4 &  4.1	  &&   2455620.054 &	  38.1 &  7.6 \\
  2452437.870 &      22.0 &  6.2 &&        2453401.097 &$-$138.2 &  5.3	  &&   2455651.019 &     167.2 &  6.1 \\
  2452438.821 &      28.1 &  6.1 &&        2453443.008 & $-$89.6 &  4.7	  &&   2455678.978 &     264.8 &  6.5 \\
  2452452.913 &      86.5 &  6.2 &&        2453445.050 & $-$79.1 &  4.9	  &&   2455701.928 &     307.6 &  5.3 \\
  2452454.847 &     101.9 &  5.9 &&        2453493.983 & 23.1 &  5.3	  &&   2455732.871 &     375.3 &  4.8 \\
  2452464.877 &     131.8 &  5.8 &&        2453578.846 & 283.0 &  5.9	  &&   2455756.780 &     347.2 &  5.3 \\
  2452472.845 &     147.1 &  5.9 &&        2453613.703 & 296.7 &  5.3	  &&   2455760.815 &     341.6 &  6.0 \\
  2452483.816 &     217.2 &  6.0 &&        2453649.645 & 254.2 &  5.9	  &&   2455803.707 &     237.0 &  5.1 \\
  2452484.767 &     217.3 &  6.5 &&        2453654.615 & 236.0 &  4.8	  &&   2455832.617 &     174.0 &  6.0 \\
  2452494.830 &     241.1 &  6.6 &&        2453788.093 &$-$228.3 &  5.4	  &&   2455862.611 &	  59.2 &  6.5 \\

\cline{1-3}\cline{1-3}\cline{5-7}\cline{5-7}\cline{9-11}\cline{9-11}
\noalign{\vskip 0.5mm}
\cline{1-3}\cline{1-3}\cline{5-7}\cline{5-7}\cline{9-11}\cline{9-11}
\noalign{\vskip 2.5mm}

%  $\star$ - CRIRES data \\
 %\makebox[0.1\textwidth][l]{absolute RV$_{CRIRES}$~=~11.849 km\,s$^{-1}$ } \par \\

% \vspace{-0.8cm}
\end{tabular}
\end{minipage}}
\end{table*}

\end{appendix}

\end{document}